\def\min{\textrm{\mbox{\tiny{min}}}}
\def\max{\textrm{\mbox{\tiny{max}}}}

\renewcommand{\sec}{\mathrm{s}}

\newcommand{\M}{\mathcal{M}}

\providecommand{\sc}[1]{\widehat{#1}}
\renewcommand{\sc}[1]{\widehat{#1}}

\newcommand{\F}{\mathcal{F}}		

\renewcommand{\csc}{$\sc{\textrm{c}}$}

\documentclass[twocolumn,twocolappendix]{aastex631}
\shorttitle{Continuous gravitational waves from Galactic neutron stars}
\shortauthors{Pagliaro, Papa, Ming et al.}
\graphicspath{{./}{figures/}}
\usepackage{amsmath}
\begin{document}

\title{Continuous gravitational waves from Galactic neutron stars: demography, detectability and prospects}

\author[0009-0008-1886-8912]{Gianluca Pagliaro}
\email{gianluca.pagliaro@aei.mpg.de}
\affiliation{Max Planck Institute for Gravitational Physics (Albert Einstein Institute), Callinstrasse 38, 30167 Hannover, Germany}
\affiliation{Leibniz Universit\"at Hannover, D-30167 Hannover, Germany}

\author[0000-0002-1007-5298]{Maria Alessandra Papa}
\email{maria.alessandra.papa@aei.mpg.de}
\affiliation{Max Planck Institute for Gravitational Physics (Albert Einstein Institute), Callinstrasse 38, 30167 Hannover, Germany}
\affiliation{Leibniz Universit\"at Hannover, D-30167 Hannover, Germany}

\author[0000-0002-2150-3235]{Jing Ming}
\email{jing.ming@aei.mpg.de}
\affiliation{Max Planck Institute for Gravitational Physics (Albert Einstein Institute), Callinstrasse 38, 30167 Hannover, Germany}
\affiliation{Leibniz Universit\"at Hannover, D-30167 Hannover, Germany}

\author[0000-0001-5258-1466]{Jianhui Lian}
\affiliation{Max Planck Institute for Astronomy, 69117 Heidelberg, Germany}
\affiliation{Department of Physics \& Astronomy, University of Utah, Salt Lake City, UT 84112, USA}

\author[0000-0002-6347-3089]{Daichi Tsuna}
\affiliation{TAPIR, Mailcode 350-17, California Institute of Technology, Pasadena, CA 91125, USA}
\affiliation{Research Center for the Early Universe (RESCEU), School of Science, The University of Tokyo, 7-3-1 Hongo, Bunkyo-ku, Tokyo 113-0033, Japan}

\author[0000-0001-7711-3677]{Claudia Maraston}
\affiliation{Institute of Cosmology and Gravitation, University of Portsmouth, Dennis Sciama Building, Portsmouth, PO1 3FX, UK}

\author[0000-0002-6325-5671]{Daniel Thomas}
\affiliation{Institute of Cosmology and Gravitation, University of Portsmouth, Dennis Sciama Building, Portsmouth, PO1 3FX, UK}



\begin{abstract}
We study the prospects for detection of continuous gravitational signals from ``normal" Galactic neutron stars, i.e. non-recycled ones. We use a synthetic population generated by evolving stellar remnants in time, according to several models.  
We consider the most recent constraints set by all-sky searches for continuous gravitational waves and use them for our detectability criteria.
We discuss detection prospects for the current and the next generation of gravitational wave detectors. We find that neutron stars whose ellipticity is solely caused by magnetic deformations cannot produce any detectable signal, not even by 3rd-generation detectors. Currently detectable sources all have $B\lesssim10^{12}$ G and deformations not solely due to the magnetic field. For these in fact we find that the larger the magnetic field is, the larger is the ellipticity required for the signal to be detectable and this ellipticity is well above the value induced by the magnetic field.
Third-generation detectors as the Einstein Telescope and Cosmic Explorer will be able to detect up to $\approx 250$ more sources than current detectors. We briefly treat the case of recycled neutron stars, with a simplified model. We find that continuous gravitational waves from these objects will likely remain elusive to detection by current detectors but should be detectable with the next generation of detectors.
\end{abstract}


\section{Introduction} 
\label{sec:intro}
Continuous gravitational waves are expected to be emitted by neutron stars that present a degree of asymmetry with respect to their rotation axis. A number of mechanisms are thought to be responsible for such deviations from perfect axisymmetric configurations. Strongly magnetised neutron stars may present a deformation proportional to their magnetic field energy \citep{ChandraFermi1953, Ferraro1954, Katz1989, Haskell2008, Mastrano2011} that in conjunction with a misalignment between rotational and magnetic field axis leads to non-axisymmetry. Accreting objects may develop \emph{mountains} due to non-axisymmetric temperature variations in the crust (thermal mountains) \citep{Bildsten1998, Ushomirsky2000} or magnetic confinement of the accreted material (magnetic mountains) \citep{Brown1998, Melatos2005, Vigelius2009, Priymak2011}. It has also been suggested that accreting neutron stars spinning rapidly enough to lead to crustal failure might eventually tend towards non-axisymmetric equilibrium geometries \citep{Giliberti2022}.

The deformations can be accommodated by elastic crustal stresses \citep{Ushomirsky2000, HaskellJones2006, Horowitz2009} or, for neutron stars with non-conventional matter composition, by elastic phases of matter in the deep core \citep{Owen2005, Haskell2008_mountains}.
Irrespective of the underlying mechanism, neutron star asymmetry is typically described by the equatorial \emph{ellipticity} $\varepsilon = {| I_{xx} - I_{yy} |}/{I_{zz}}$, where $I_{ii}$ is the moment of inertia referred to the $i$ axis ($i = x, y, z$) and where $z$ is aligned with the spin axis.

Continuous gravitational waves emitted by non-axisymmetric spinning neutron stars are nearly-monochromatic signals that are practically ``ON" all the time. These signals are profoundly different from the gravitational wave signals detected so far, which all come from compact binary coalescences -- catastrophic events, leading to major transformations of the emitting system, and lasting of the order of seconds. The amplitude of continuous waves is several orders of magnitude smaller than that of coalescence signals and this major drawback is partly compensated by the fact that they are long-lived, and in principle one can build up signal-to-noise ratio by integrating the data over time. This unfortunately comes with huge computational costs.

The least computationally expensive continuous wave searches are the so called \emph{targeted} searches, where one targets a known object such as a pulsar, with sky-position and phase parameters (frequency and its time derivatives) known from electromagnetic observations \citep{LIGOScientific:2021hvc,LIGOScientific:2021quq,LIGOScientific:2021ozr,Ashok:2021fnj,Nieder:2020yqy,Rajbhandari:2021pgc}. 

If the sky position of the object is known, but no information is available on the phase parameters, one can set up a \emph{directed} search, and explore the free parameter space. This may be constrained by information on the object, such as its age \citep{Owen:2022mvu,2022PhRvD.106f2002A,Ming:2021xtz,LIGOScientific:2021mwx,Zhang:2020rph}. These searches are directed towards objects like supernova remnants, LMXBs or promising regions in the sky, and their computational cost is considerably higher than that of the searches for known pulsars. 

At the top of the computational cost ladder are the all-sky surveys, where there are no specific targets, and instead the aim is to detect a signal from a previously unidentified object. In this case assumptions on the expected signal population define the surveyed parameter space. Extensive searches are carried out (for a sample of recent results see \citet{2023ApJ...952...55S, 2022PhRvD.106j2008A,Covas:2022rfg,2023PhRvX..13b1020D, KAGRA:2021una,Dergachev:2021ujz}), which have translated the no-detection results into constraints on the physical parameters of the sub-population investigated so far.

What fraction of the Galactic neutron star population is actually probed by the searches? 
\cite{Reed2021} found that while recent O2 data all-sky searches probe ellipticities below $10^{-6}$ for nearby objects, overall they rule out ellipticities below $10^{-5}$ for only $\approx 1.6\%$ of all Galactic neutron stars.

The goal of this paper is to contribute to answering the question of the significance of the sample of objects probed by current searches, by factoring-in the astrophysical parameters relevant for the emission and the detection of continuous waves, their evolution time, and adding a detectability assessment to the discussion.

We use a population synthesis approach and generate a dataset of $\approx 4.5 \times 10^{8}$ isolated non-axisymmetric normal (non-recycled)  neutron stars, starting from an initial distribution of neutron star progenitors, dynamically evolving them throughout the Galaxy under the influence of the Galactic potential. 

We consider different models that correspond to different combinations of the astrophysical priors that determine the spin evolution. We obtain  various present-time populations, from which we study the characteristics of objects detectable by present and future detectors and their parameters. 

This paper is organised as follows: in Section \ref{sec:population} we present the details of the synthetic population; in Section \ref{sec:models} the astrophysical priors that define the evolution models summarised in Section \ref{sec:summary}. In \ref{sec:spin_evol} and \ref{sec:detect} we present and discuss our results. Recycled neutron stars are treated in Section \ref{sec:recycled}. We draw our conclusions in Section \ref{sec:conclusions}.

\section{The Synthetic neutron star population: distribution in space} 
\label{sec:population}
\begin{figure}
	\centering
    \caption{Surface density distribution of our synthetic neutron star population on the sky. The dashed black line indicates the Galactic plane and the magenta star marks the Galactic center.} 
	\includegraphics[width=\columnwidth]{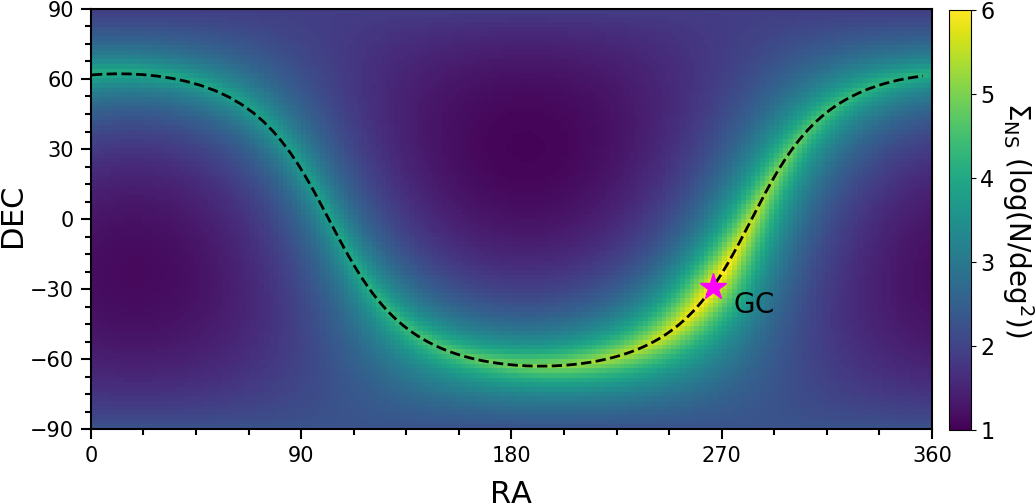}
	\label{sky-map}
\end{figure} 

Neutron stars are the stellar evolution remnants of massive stars. The lowest mass star likely to generate a neutron star is around 8~$M_{\odot}$. Very high mass stars are more likely to produce black holes than neutron stars, and the highest mass star that still generates a neutron star depends on the specifics of the object's evolution, such as its mass loss and rotation history. Indicatively we take the highest mass limit to be 40~$M_{\odot}$ (following \cite{1993ApJ...416L..49R}).

We generate a synthetic population of stars, identify those in the mass range [8, 40]~$M_{\odot}$ and take them as the progenitors of neutron stars with mass $M=1.4~M_{\odot}$ \citep{Renzini:1993xh,Maraston:1998dy}. 

The star population is based on an initial stellar mass function (IMF), a simplified stellar spatial density model and the formation-rate history in the Milky Way.
 
For the IMF we follow \citealt{kroupa2001}. 
 
According to the stellar population models of \citet{maraston2005,Maraston:1998dy}, for a Kroupa IMF, for instantaneous\footnote{Each star forms ``instantaneously'', i.e. in a single burst.} stellar populations older than $\approx$ 30 Myr (i.e. the lifetime of an 8 solar mass star), we expect approximately 1\% of the mass of
all stars to be neutron stars. For simplicity, we assume the same mass fraction in neutron stars for populations of all ages older than ~4 Myr (i.e. the lifetime of a 40M$_\odot$ star, which is the mass threshold above which the remnant will be a black-hole). In these calculations, the only (mild) dependence on metallicity is the one of the turnoff mass and its lifetime (see \cite{maraston2005}).
 
The density model focuses on the most massive stellar component of our Galaxy, the thin disc \citep{licquia2016}, for which a standard exponential disc structure is assumed with scale length of 2.6~kpc and scale height of 0.3~kpc \citep{bland2016}. 
We assume a spatially-invariant star formation history (SFH) within the thin disc for simplicity, although a radially-varying SFH is suggested in studies based on detailed stellar chemical abundances \citep[e.g.,][]{chiappini2001,lian2020a,lian2020b}.  

We assume an exponentially decreasing gas accretion history with e-folding time of 10~Gyr, which is representative of the solar radius \citep{chiappini2001}, and the Kennicutt-Schmidt star formation law \citep{kennicutt1998}. We set an initial gas accretion rate of 0.01~${\rm M_{\odot}yr^{-1}}$. This results in a present-day stellar mass density of 0.050~${\rm M_{\odot}pc^{-3}}$, which is close to observational results of 0.040-0.043~${\rm M_{\odot}pc^{-3}}$ in the solar neighbourhood \citep{flynn2006,mckee2015,bovy2017}. 
We also calculate the chemical enrichment history given the adopted gas accretion history using the chemical evolution model of \citet{lian2018a,lian2020a}.

By sampling the SFH we obtain a synthetic catalog of neutron star progenitors that contains 3D spatial position, age, metallicity for $\sim4.5\times 10^{8}$ objects which are broadly in line with other estimates in the literature \citep{Sartore:2009wn,Diehl:2006cf}.

Figure~\ref{sky-map} shows the density distribution of neutron star-progenitors in our synthetic catalog for all ages and distances on the sky. It is plotted in equatorial coordinates assuming an Earth position at 8.2~Kpc in Galactocentric radius and 27 pc above the disc plane \citep{bland2016}.
As expected, the overall density closely follows that of stars, peaking in the direction of the Galactic center, decreasing mildly in the anti-Galactic direction along the disc plane, and dropping dramatically when moving away from the disc plane. The age is independent of sky position.

We associate with each neutron star progenitor a velocity based on disc stars observed today. This is obtained using 3D velocity measurement of $\sim$300,000 disc stars in the APOGEE survey \citep{majewski2017} provided by the astroNN catalog \citep{mackereth2018,leung2019b}, which are calculated using spectroscopic observations from APOGEE and astrometric observations from Gaia.

We evolve the position of each neutron star since birth using  the procedures described in \citep{Tsuna_2018}. We account for the initial velocity of progenitors, natal kicks and the Galactic gravitational potential. 

Recent neutron star population synthesis studies (e.g. \citet{Vigna-Gomez:2018dza}) adopt two types of kicks depending on the supernova type. One is the conventional Maxwell distribution with $\sigma=265$ km/s for core-collapse supernovae, and the other with $\sigma = 30$ km/s for electron-capture supernova.  The kick velocity distribution we use is a combination of these two kick distributions.
We assume a 75\% core-collapse supernova population and 25\% electron-capture supernova population reflecting the fact that this paper focuses on isolated neutron stars.

For each neutron star, we assign the kick velocity from the combined distribution and we uniformly randomly assign a direction for that kick velocity. Combined with the initial velocity from its progenitor, the  trajectory is evolved for the duration of its age. Thus, the velocity and  spatial  position of each neutron star  at the present time is determined.

\section{Spin-down model and astrophysical priors} 
\label{sec:models}

A magnetised, non-axisymmetric object looses rotational kinetic energy due to the emission of electromagnetic and gravitational waves. In the presence of an external dipolar magnetic field $B(t)$ and an equatorial ellipticity $\varepsilon$, the star's spin frequency $\nu$ evolves as
\begin{equation}
\dot{\nu} = -\frac{32 \pi^3 R^6}{3 I c^3 \mu_0} \left(B(t) \sin \chi \right)^2 \ \nu^3 - \frac{512 \pi^4 G I}{5 c^5} \varepsilon^2 \nu^5.
\label{eq:combsd}
\end{equation}
We use this equation -- valid in SI units and first introduced by \cite{Ostriker1969} -- to evolve the spin of the star from birth to current time. We will use the subscript ``0" to refer to quantities at birth.

In Equation~\ref{eq:combsd} it is assumed that the spin axis coincides with one of the star's principal moment of inertia (no precession), and that the magnetic dipole moment is misaligned with respect to the rotation axis by an angle $\chi$. $R$ is the radius of the star, and $I$ is the moment of inertia about the rotation axis. We fix the radius and the moment of inertia to the fiducial values $R=12$ km and $I=10^{38}$ kg m$^2$. For simplicity we also assume $\sin \chi = 1$, thus considering every object to be an orthogonal rotator, also neglecting any spin-axis evolution in time. This, as shown by \cite{Bonazzola1996}, implies that gravitational waves will be composed of a single harmonic at frequency $f_{GW} = 2 \nu$. 

In Section~\ref{sec:agnosticModel} we consider very broad models, encompassing the most common scenarios and parameter ranges. In this respect these are ``Partly Agnostic Models" but for ease of notation we will refer to them as ``Agnostic Models" in the rest of the paper. In Section~\ref{sec:empirical} we instead follow a specific ``Empirical" model, based on the works of \citep{Popov2010, Vigano2013, Gullon2015}.

\subsection{Agnostic, A Models}  \label{sec:agnosticModel}

These models are defined by a single distribution of neutron stars in the sky and a single magnetic field distribution, two distributions for birth spin-frequencies and two distributions for the ellipticity. Hence, all in all, we have four different A models. All the distributions are described below.

\subsubsection{Magnetic fields}\label{sec:magfield}

Emission from isolated spinning neutron stars has been observed in the form of radio, X-ray and $\gamma$-ray pulses, as well as non-pulsed radiation. The electromagnetic activity in neutron stars is intimately related to their magnetic field, whose magnitude can significantly vary depending on the neutron star-type. Non-recycled isolated neutron stars comprise radio-quiet central compact objects with inferred external magnetic fields of the order of $B = 10^{10-11} \, G$, radio pulsars with $B = 10^{12-13} \, G$ and magnetars with $B = 10^{14-15} \, G$ \citep{Kaspi2016,Teruaki2019}.
We hence consider a broad and agnostic distribution of magnetic field values: 
\begin{equation}
\label{eq:magfieldprior}
p(\log B/[G] )=\mathcal{U}(10,15 )
\end{equation}
where $\mathcal{U}(x_{\textrm{min}},x_{\textrm{max}})$ indicates a uniform distribution between $x_{\textrm{min}}$ and $x_{\textrm{max}}$.
We assume constant (time independent) magnetic fields.

\subsubsection{Ellipticity} \label{sec:ellip}

The magnetic field of a neutron star causes deformations of its shape from perfect sphericity. 
Magnetic stresses are generated by the interaction between the conducting neutron star interior and the internal fields, producing deformations that scale linearly with the magnetic field energy. 
Assuming a mixed poloidal-toroidal geometry for the internal magnetic field, \cite{Mastrano2011} provide a relation that, if rescaled to our fiducial value for the radius, assuming the mass of the neutron star to be $1.4 \, M_{\odot}$ reads
\begin{equation}
\varepsilon(B, \Lambda) = 12.985 \times 10^{-6} \left( \frac{B}{5 \times 10^{14}  \mathrm{G}} \right)^2 \times \left( 1 - \frac{0.385}{\Lambda} \right),
\label{eq:mastrano}
\end{equation}
where $\Lambda$ is the fraction of the magnetic energy stored in the poloidal component over the total magnetic energy of the star. In order to obtain it, the authors impose continuity at the surface of the star between the internal mixed field and an external dipole (e.g. the toroidal component vanishes at the surface). This is how the deformation is connected with the {\it{external}} magnetic field, that drives the spin evolution. However, it must be understood that it is the {\it{internal}} field that is responsible in the deformation, not the external one \citep{AnderssonBook}, and that unfortunately there is no direct measurement of the former. 

The range of variability for $\Lambda$ is largely unknown. The consensus is that neither pure poloidal nor pure toroidal configurations are stable \citep{Markey1973, Tayler1973, Wright1973, Braithwaite2007, Kiuchi2008, Lander2011, Ciolfi2011}, but whether the magnetic field energy is dominated by one of the two components is a matter of debate. 
In favour of configurations where the poloidal energy component is dominant we point to the works by \cite{Ciolfi2010}, \cite{Lasky2012}, \cite{Lander2009} and \cite{Lander2021}, while \cite{Braithwaite2009}, \cite{akgun2013} and \cite{Ciolfi2013}, have argued in favour of a dominant toroidal component. 

The last factor on the RHS of Equation~\ref{eq:mastrano} equals to 0 when $\Lambda^{\prime} = 0.385$, it is positive (negative) if $\Lambda > \Lambda^{\prime}$ ($\Lambda < \Lambda^{\prime}$). A positive ellipticity means an oblate shape, while a negative ellipticity means a prolate one. 

We consider two scenarios:
\begin{description}

    \item[Model 1]
The magnetic field is the source of the deformation. For each synthetic object, with an assigned value for the external magnetic field $B$, we consider two values of $\Lambda$ as follows:
\begin{itemize}
    \item $\Lambda = 0.9$, consistent with \citeauthor{Ciolfi2010, Lander2021} and roughly with \citeauthor{Lander2009}.  We note that \cite{Lasky2012} find a stable configuration for $\Lambda = 0.65$. The resulting ellipticity is however very close to that of the $\Lambda = 0.9$ configuration, making the two equivalent for the purpose of this study. 
    \item $\Lambda = 0.1$, consistent with where results by \citeauthor{Braithwaite2009,akgun2013,Ciolfi2013} overlap.
\end{itemize}
For each $B$ and $\Lambda$ we take the ellipticity to be 
\begin{equation}
\label{eq:scenarioOneEpsilon}
\varepsilon={\textrm{min}}[\varepsilon(B, \Lambda),\varepsilon_{max}],
\end{equation}
with $\varepsilon(B, \Lambda)$ from Equation~\ref{eq:mastrano}. Based on estimates of the maximum possible ellipticity sustainable by a neutron star \citep{Ushomirsky2000, HaskellJones2006, Horowitz2009,Gittins2021-1, Gittins2021-2, 2022MNRAS.517.5610M}, we set $\varepsilon_{max} = 10^{-5}$.

\item[Model 2] \label{sec:scen2}
We do not specify the origin of the ellipticity, but we draw its value from a log-uniform distribution where the lower bound is due to the magnetic field. For each synthetic object, with an assigned value for the external magnetic field $B$, we draw the ellipticity from the following distribution 
\begin{equation}
\label{eq:epsilonScenarioTwoPrior}
p(\log \varepsilon )=\mathcal{U}(\log\varepsilon_{min},\log\varepsilon_{max})
\end{equation}
and
\begin{equation}
\label{eq:epsilonScenarioTwoLimits}
        \begin{cases}
      \varepsilon_{min}=\varepsilon(B, \Lambda=0.9)\\
      \varepsilon_{max} = 10^{-5}.
        \end{cases}
\end{equation}
We choose $\Lambda = 0.9$ since it gives, with equal magnetic fields, smaller ellipticities.

\end{description}

Two clarifications are needed. First, the ellipticity given originally by \citeauthor{Mastrano2011} is not the equatorial ellipticity, but rather the \emph{oblateness}. Since we are considering orthogonal rotators these two coincide, and in the case of prolate objects, we simply take the absolute value of the ellipticity given by \ref{eq:mastrano}. 
Second, we ignore the stability and secular evolution of oblate/prolate neutron stars with a fluid interior that exerts a centrifugal pressure on a solid crust. Such pressure would tend to create a centrifugal bulge, thus a further change in shape (and likely in ellipticity) that might not be supported by the crust. Also, spinning oblate objects with a fluid interior are subject to a secular evolution different than prolate ones, with the former aligning their symmetry axis with the spin axis, and the latter tending to orthogonal rotators configurations \citep{Cutler2000}.

\subsubsection{Birth spin frequency}
\label{sec:birthspin}

The birth of a neutron star follows the hydrodynamical instability of the progenitor and its subsequent gravitational collapse(s). 
The remnant is more compact than the parent core so the newborn neutron star possesses a much faster spin than the parent core. For a review on the topic see \cite{Janka2001}. 

Theoretical efforts have been devoted to understanding the phenomenology of neutron star birth and the spin properties of newborn neutron stars \citep{Heger2000, Heger2004, Heger2005, Ott2006, Camelio2016, Linhao2019}. Several authors have tried to extract information on the spin frequency at birth based on observational evidence \citep{FaucherGiguere2006, Perna2008, Popov2012}. In general, there is no consensus on the spin frequency distribution of newborn neutron stars.

A reasonable maximum spin frequency at birth, is given by the Keplerian break-up limit, at which the spin is fast enough that the centrifugal force overcomes the star's own gravitational energy and breaks it apart. \cite{HaskellZdunik2018} estimate an equation-of-state-independent  lower limit for the Keplerian break-up frequency at $\approx 1200 \, \textrm{Hz}$. This value, however, applies only to matter and configurations of mature neutron stars. On the other hand, \cite{Camelio2016} show that the shedding-mass limit increases -- and eventually saturates -- during the first few tens of seconds after the core bounce. Applying such limit during this phase of the proto-neutron star leads to spin frequencies generally below $300 \, \textrm{Hz}$. In this paper we will consider both estimates of the maximum birth spin frequency.
 
How slow can a neutron star spin at birth? The parameter that mostly affects the spin frequency at birth is the initial angular momentum of the post bounce iron core \citep{Ott2006, Camelio2016}, which is largely unknown and poorly constrained. 

Two main strategies exist to predict neutron star birth spin distributions that resort to observational constraints: i) consider a set of pulsars with well known phase parameters and evolve the spin frequency back in time to their birth \citep{Popov2012}  ii) consider a synthetic population of neutron stars  such that, when evolved to the present era,  the observed spin distribution for pulsars is recovered \citep{FaucherGiguere2006}. In addition, \cite{Perna2008} propose a method based on the X-ray luminosities of a relatively large sample of supernovae to constrain the initial spin frequency of the newborn neutron star. In all the studies above the slowest newborn neutron stars born have spin periods of $few$ hundreds of milliseconds. In light of this we set our lower bound to $2 \, \textrm{Hz}$ that corresponds to a spin period of $500 \, ms$.

We consider two log-uniform priors for the birth spin frequency:
\begin{itemize}
\item low:	$p(\log \nu_0/[\textrm{Hz}])=\mathcal{U}(\log 2, \log 300)$
\item high: $p(\log \nu_0/[\textrm{Hz}])=\mathcal{U}(\log 2, \log 1200)$.
\end{itemize}

\subsection{Empirical, E Models} 
\label{sec:empirical}

Our "Empirical" models are based on the studies of \cite{Popov2010, Vigano2013, Gullon2015}.
These works assume electromagnetic spin-down only, and tune the magnetic field and birth spin frequency distributions, in order to match the joint distributions of the observed population of magnetars, normal radio pulsars and thermally emitting neutron stars. They also track the magnetic field decay through magneto-thermal evolution codes. Here we consider models for the magnetic field and the birth spin frequency based on their results, as described below. For the ellipticity we consider the distributions described in Section~\ref{sec:ellip}, but with the magnetic field values of this model.

Summarizing, the E models are defined by a single distribution of neutron stars in the sky and a single magnetic field distribution, two distributions for birth spin-periods and two distributions for the ellipticity. Hence, all in all, we have four different E models. All the distributions are described below.

\subsubsection{Magnetic field}
\label{sec:S_B}
For the distribution of magnetic fields at birth $B_0$, loosely following \cite{Gullon2015}, we draw values from a truncated log-normal distribution:
\begin{equation}
\label{eq:B0Prior}
    p(\log B_0/[G] )=
    \begin{cases}
      \mathcal{N}(13,0.8) & \text{if} \ B_0 \leq 10^{15} \, {\textrm{G}} \\
      0 & \text{otherwise}
    \end{cases}
\end{equation}
with  $\mathcal{N}(\mu,\sigma)$ indicating a normal distribution with mean $\mu$ and standard deviation $\sigma$.

In order to account for magnetic field decay, following \cite{Chattopadhyay2020,Cieslar2020}, we implement the following time dependence
\begin{equation}
\label{eq:decay}
B(t)=(B_0 - B_{min})e^{-t/\tau_B} + B_{min},
\end{equation}
where $\tau_B$ is the decay timescale and $B_{min}$ is a lower limit for $B$. Equation \ref{eq:decay} essentially describes a field that decays exponentially when $t \lesssim \tau_B$ and then saturates to values of the same order as $B_{min}$ for $t > few \cdot \tau_B$. This implementation wants to mimic what found in magneto-thermal simulations where the magnetic field decays significantly in the first $10^5-10^6$ yr to then slow down and, from a few million years after birth, remains almost constant \citep{Popov2010}. We hence choose $\tau_B = 10^6$ yr and set $B_{min}$ to the value taken by the magnetic field at $t = 10 \cdot \tau_B$ as if the decay were purely exponential (i.e. $B_{min} = B_0 \cdot e^{-10}$). 

\subsubsection{Birth spin frequency}
\label{sec:S_P0}

The authors of the studies mentioned at the beginning of this section all reason in terms of spin period rather than spin frequency. We thus draw  spin periods $P_0$ for the newborn neutron stars of our synthetic population, and then transform these into spin frequencies through the relation $\nu_0 = \frac{1}{P_0}$. 
We consider two models:
\begin{itemize}
\item norm:  $p(P_0/[ms])=\mathcal{N}(300,200)$. This is consistent with \citep{Popov2010,Vigano2013,Gullon2015}.
\item unif: $p(P_0/[ms])=\mathcal{U}(0.8, 500)$. This is justified by the fact that \cite{Gullon2015} argue that the overall population statistics is not very sensitive to the initial spin distribution, and find that by slightly re-adjusting the rest of the parameters, any nearly uniform distribution in the range $0 < P_0/[ms] < 500$ reproduces just as well the present-day spin distribution. To some extent the same conclusion is drawn by \citet{Gonthier2004}, who consider the radio-pulsar population only, and by \citet{Gullon2014}, who additionally consider X-ray thermally emitting pulsars. The minimum value of $0.8 \, {\textrm{ms}}=1250 {\textrm{\textrm{Hz}}}$ roughly corresponds to the Keplerian break-up frequency mentioned in \ref{sec:birthspin}.
\end{itemize}

\subsection{Models summary} \label{sec:summary}

In total we have eight models: four are the agnostic A models of Section~\ref{sec:agnosticModel} and four are the empirical E models of Section~\ref{sec:empirical}. The four combinations come from two models for $\varepsilon$ -- 1 and 2\footnote{Model 1 ellipticity models actually consider both values of $\Lambda$ as explained in section \ref{sec:ellip}. For each such model every realisation is performed twice, once per value of $\Lambda$.} -- and two models for $f_0$ or $P_0$ (high/low and unit/norm for A and E respectively). So all in all we have: $\textrm{A1}_{\textrm{low}}$, $\textrm{A1}_{\textrm{high}}$, $\textrm{A2}_{\textrm{low}}$, $\textrm{A2}_{\textrm{high}}$, $\textrm{E1}_{\textrm{unif}}$, $\textrm{E1}_{\textrm{norm}}$, $\textrm{E2}_{\textrm{unif}}$ and $\textrm{E2}_{\textrm{norm}}$. A summary of the model parameters is given in Table \ref{tab:models}.

\centerwidetable
\begin{deluxetable*}{lccc}
\label{tab:models}
\tablecaption{Models summary.}
\tablehead{\colhead{Model name} & \colhead{Magnetic field} & \colhead{Birth spin frequency} & \colhead{Ellipticity} \\ 
\colhead{} & \colhead{[G]} & \colhead{[\textrm{Hz}]} & \colhead{}}

\startdata
$\textrm{A1}_{\textrm{low}}$ & log-uniform: $10^{10} \leq B \leq 10^{15}$ & log-uniform: $2 \leq \nu_0 \leq 300$ & model 1  \\
$\textrm{A1}_{\textrm{high}}$ & " & log-uniform: $2 \leq \nu_0 \leq 1200$ & " \\
$\textrm{A2}_{\textrm{low}}$ & " & log-uniform: $2 \leq \nu_0 \leq 300$ & model 2 \\
$\textrm{A2}_{\textrm{high}}$ & " & log-uniform: $2 \leq \nu_0 \leq 1200$ & " \\
$\textrm{E1}_{\textrm{norm}}$ & log-normal:\tablenotemark{\small \textup{a}} $\mu_{\log_{10}B_0}=13$, $\sigma_{\log_{10}B_0}=0.8$  & See Section \ref{sec:empirical} & model 1 \\
$\textrm{E1}_{\textrm{unif}}$ & "  & See Section \ref{sec:empirical} & " \\
$\textrm{E2}_{\textrm{norm}}$ & "  & See Section \ref{sec:empirical} & model 2 \\
$\textrm{E2}_{\textrm{norm}}$ & "  & See Section \ref{sec:empirical} & " \\
\enddata
\tablenotetext{}{\textbf{Notes.} The ditto mark means ``same as above".}
\tablenotetext{\textrm{a}}{Cut-off at $10^{15} \, G$.}
\end{deluxetable*}

\section{Spin evolution}
\label{sec:spin_evol}

Every newborn neutron star has a position, age, and an initial velocity; as described in Section \ref{sec:population} its position is evolved in time yielding a present-day one. The present-time $\nu - \dot{\nu}$ distribution is determined by values drawn from the different distributions of $B_0$ (or $B$ in the case of static magnetic fields), $\varepsilon$ and $\nu_0$, depending on the model, and Equation~\ref{eq:combsd}. We thus obtain a distinct population of neutron stars for each model adopted. 

We integrate the spin-down Equation~\ref{eq:combsd} approximating the spin evolution to be solely determined by magneto-dipole emission - in limiting cases - since then the present-time spin frequency is attainable analytically (see for example section II of \citet{Wade2012}). In particular, setting $\sin \chi = 1$ we can re-write the spin-down Equation~\ref{eq:combsd} as 
\begin{equation}
\label{eq:dotnu}
\begin{cases}
 \dot{\nu} = \gamma_{dip} \nu^3 + \gamma_{GW} \nu^5\\
\gamma_{dip} = - \frac{32 \pi^3 R^6}{3 I c^3 \mu_0} B^2 , \quad \gamma_{GW} = - \frac{512 \pi^4 G I}{5 c^5} \varepsilon^2 .
\end{cases}
\end{equation}
In the case of a stationary magnetic field (models $A$), the spin-down is $\approx$ purely magneto-dipolar if
\begin{equation}
\label{eq:emsdcond}
\frac{\gamma_{GW} \nu^5}{\gamma_{dip} \nu^3} < \frac{1}{100} \: \Rightarrow \: \gamma := {{\gamma_{GW}}\over{\gamma_{dip}}} <  \frac{10^{-2} }{\nu^2}
\end{equation}
Since the RHS of Equation~\ref{eq:emsdcond} is minimised at higher frequencies, we choose our condition for pure magneto-dipole spin-down to be
\begin{equation}
\label{eq:upperGamma}
\gamma < 10^{-8} \, s^2
\end{equation}
which is what we get from Equation~\ref{eq:emsdcond} if we set $\nu = 1000 \, \textrm{Hz}$, close to the highest frequency of our models.

For the models with magnetic field decay (models E), the condition for pure magneto-dipolar emission has to be satisfied at present time (lowest $B$).
The analytical relation that gives present time spin frequency in this case is given by \cite{Chattopadhyay2020} in Equation~(6) of their work.

The gravitational-wave-dominated spin-down limit cannot occur for models A, while it is extremely improbable and in practise never realised for models E since it requires ellipticities $\approx 10^{-5}$ and magnetic fields about 5 sigmas away from their mean value.

In the $\gamma$ ranges outside of Equation~\ref{eq:upperGamma} we integrate numerically following different procedures based on the model considered. These are explained in detail in appendix \ref{sec:num_integ}. 
\begingroup
\setlength{\tabcolsep}{10pt}
\begin{deluxetable}{lcl}
\label{tab:inband}
\tablecaption{Number of sources \emph{in-band} and relative percentage with respect to the Galactic neutron star population for current (${N}_{f_{GW}>20\textrm{Hz}}$) and future detectors (${N}_{f_{GW}>5\textrm{Hz}}$) computed as the average over the 100 realisations performed. 
}
\tablehead{ 
\colhead{Model}&\colhead{ ${N}_{f_{GW}>20\textrm{Hz}}$}&\colhead{ ${N}_{f_{GW}>5\textrm{Hz}}$ }
} 
\startdata
 $\textrm{A1}_{\textrm{low}}$&$4.9 \times 10^{5} (0.14\%)$&$9.9 \times 10^{6} (2.9\%)$  \\
  $\textrm{A1}_{\textrm{high}}$&$5.6 \times 10^{5} (0.16\%)$&$10.2 \times 10^{6} (3\%)$ \\
  $\textrm{A2}_{\textrm{low}}$&$4.8 \times 10^{5} (0.14\%)$&$9.9 \times 10^{6} (2.9\%)$  \\
  $\textrm{A2}_{\textrm{high}}$&$5.5 \times 10^{5} (0.16\%)$&$10.2 \times 10^{6} (3\%)$ \\
  $\textrm{E1}_{\textrm{norm}}$&$7.4 \times 10^{5} (0.22\%)$&$23.2 \times 10^{6} (6.9\%)$  \\
  $\textrm{E1}_{\textrm{unif}}$&$1.6 \times 10^{6} (0.47\%)$&$29.1 \times 10^{6} (8.7\%)$  \\
  $\textrm{E2}_{\textrm{norm}}$&$6.7 \times 10^{5} (0.20\%)$&$23.1 \times 10^{6} (6.9\%)$  \\
  $\textrm{E2}_{\textrm{unif}}$&$1.4 \times 10^{6} (0.43\%)$&$29.1 \times 10^{6} (8.7\%)$ \\
\enddata
\tablenotetext{}{\textbf{Notes.} For models 1 we show only the results for $\Lambda = 0.1$. The results for $\Lambda = 0.9$ are very similar. During the course of its life, a neutron star may drift outside of the Galaxy -- with a kick velocity of $300 \textrm{km} \textrm{s}^{-1}$, a $10^{10}$ yr old neutron star can drift by 300 kpc. Out of the $\approx 4.5 \times 10^8$ in the initial population, we now find ${N} \approx 3.5\times10^8$within 50 kpc of the Galactic center, which is a generous estimate of the horizon distance for continuous waves in this frequency range for the next decade of observations. This is what we will consider as our population.}
\end{deluxetable}
\endgroup
For each model we perform 100 realisations of the synthetic population. The frequency evolution of each object is calculated on the ATLAS cluster at AEI Hannover.

Table \ref{tab:inband} shows the number of sources whose gravitational wave frequency -- rotation frequency $\times$ 2 -- lies in the band of ground-based gravitational wave detectors. We see that less than 1\% of the population of considered neutron stars falls in the band of the current detectors, but if the lower frequency is pushed down by only 15 Hz, the number of sources increases 20-fold. A band starting at 5 Hz is plausible for the next generation of gravitational wave detectors.

The nearest neutron star is found on average at a distance of $11\pm3$ pc, consistently with what can be estimated based on general arguments (see for example \cite{Dergachev:2020fli}), but it increases to $94\pm35$ pc if we only consider neutron stars spinning ``in band".

\section{Detectability} \label{sec:detect}

With present-time spin frequency at our disposal we can evaluate both the instantaneous intrinsic spin-down through \ref{eq:combsd} and the dimensionless strain amplitude $h_0$ resulting from each synthetic neutron star 
\begin{equation}
\label{eq:h0}
h_0 = \frac{4 \pi^2 G I}{c^4 d}f_{GW}^2 \varepsilon,
\end{equation}
where $d$ is the neutron star's distance from Earth.

We consider the following three recent all-sky searches: 
\renewcommand\labelitemi{{\boldmath$\cdot$}}
\begin{itemize}
\item \citep{Dergachev:2021ujz,2023PhRvX..13b1020D}  based on Advanced LIGO O2 and O3 data, respectively (sometimes referred to as the \emph{Falcon searches})
\item \citep{2022PhRvD.106j2008A} based on Advanced LIGO and Advanced Virgo O3 data\footnote{We consider all pipelines except the \emph{SOAP} pipeline which is the least sensitive.}
\item (\cite{2023ApJ...952...55S}) based on Advanced LIGO O3 data.
\end{itemize}

A neutron star is considered to be detectable by a certain search if its gravitational-wave signal frequency and frequency-derivative fall within the parameter space covered by that search and if it has an amplitude $h_0$ greater or equal to the upper limit set by the search at that frequency. We say that an object is \emph{currently detectable} if it is detectable by at least one of the searches considered above.

Table \ref{tab:detect} shows that the percentage of in-band sources that give rise to signals that could be currently detectable is very small. With such a small number of detectable signals it is hard to extract a reliable expected value for the closest detectable star. For model $\textrm{A2}_{\textrm{high}}$ - the one with the highest statistics - we compute the average distance of the closest and of the farthest detectable neutron star. We find $d_{close} = 1.05\pm1.08\,\textrm{kpc}$, $d_{far} = 5.71\pm3.01\,\textrm{kpc}$, with the 70\% of the closest detectable within 1.5 kpc of Earth.

\subsection{Magnetically deformed Neutron Stars} \label{sec:mag_def_NS}

Let us consider first A1 and E1 models, where the ellipticity stems only from the magnetic field. We find that they cannot produce signals detectable by current nor by the future detectors considered in Sec.~\ref{sec:thirdgendet}, irrespective of the value of $\Lambda$. In fact the loudest signals from these models have amplitudes around $10^{-29} - 10^{-28}$ which is three to four orders of magnitude smaller than current upper limits and more than 50 times smaller than the estimated sensitivity to continuous gravitational waves of 3rd-generation detectors (this can be seen in the summary-figure, Figure~\ref{fig:2dhists_scen1}, at the end of the paper).

\begin{figure*}
\caption{$f_{GW}-h_0$ of neutron star populations. The color encodes the ellipticity. The shaded region indicates out-of-band frequencies. In the $\textrm{A1}$ models the ellipticity only stems from the magnetic field. In the $\textrm{A2}$ models the population is endowed with ellipticities drawn from a log-uniform distribution. We see that the highest-ellipticity objects in model A1 spin out of the band that is useful for detection, and those that remain in the useful band can do so because they have a small magnetic field and hence a low ellipticity. Conversely, in model A2, the ellipticity is not solely dependent on the magnetic field, and hence we have low magnetic field objects that remain in band that are endowed with a high ellipticity.
}
\centering
\includegraphics[width=\textwidth]{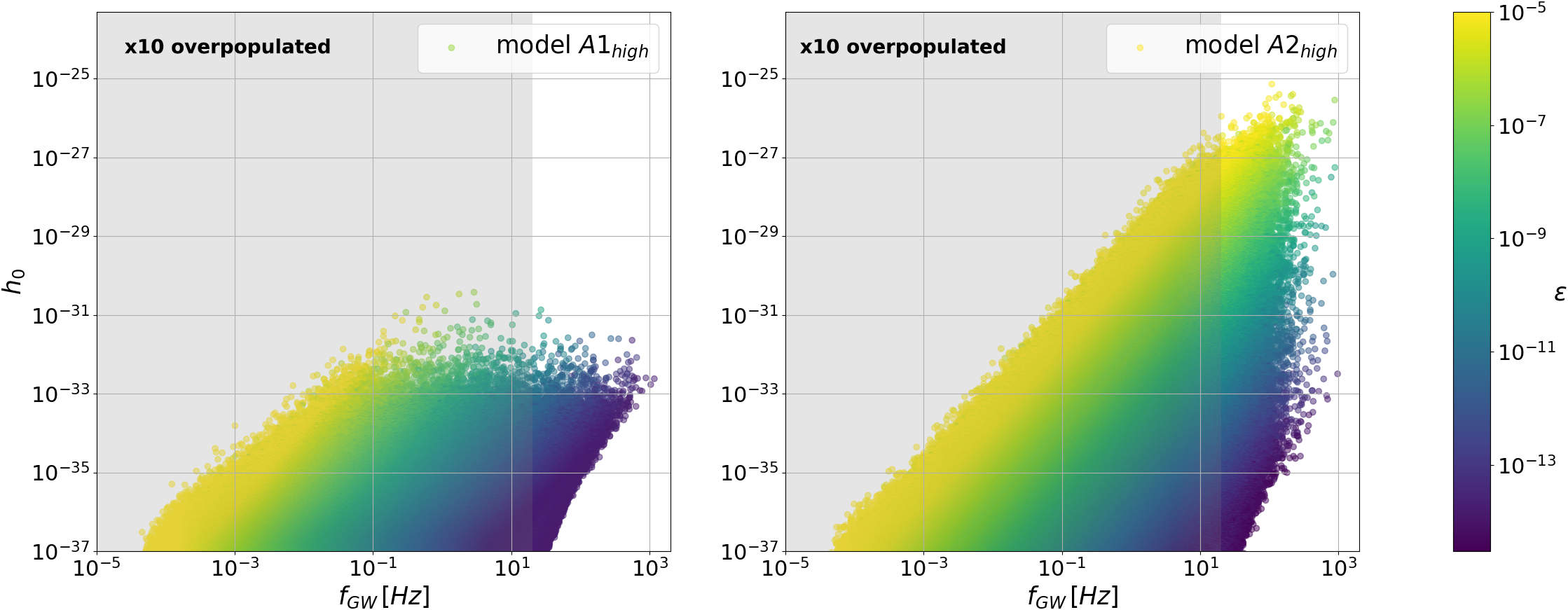}
\label{fig:f_h0_eps}
\end{figure*}

The reason why the loudest signals from these models are still very weak is that sources with a high value of ellipticity, must have a very high magnetic field and consequently a very fast spin-down that quickly pulls them to very small frequencies, and out of the instruments' band. Additionally, since $h_0\propto f_{GW}^2$ a smaller frequency means a quadratically smaller $h_0$ (see Equation~\ref{eq:h0}), so as the star spins down, the amplitude of the gravitational wave decreases. The only objects whose frequency remains high are those with small magnetic fields and hence the tenuously deformed ones (see Figure~\ref{fig:f_h0_eps}).

\subsection{Currently detectable sources} \label{sec:curr_det_sources}

\begin{deluxetable}{clccccccc}
\label{tab:detect}
\tablecaption{Expected number of currently detectable neutron stars, computed as the average $\pm ~1\sigma$ over the 100 realisations performed.}
\tablehead{ & \colhead{Model} & & & \colhead{$\overline{n}$} & & & $\%$ of \emph{in-band} & } 
\startdata
 & $\textrm{A2}_{\textrm{low}}$ & & & $1.4 \pm 1.16$ & & & $0.0003$ & \\
 & $\textrm{A2}_{\textrm{high}}$ & & & $3.62 \pm 1.91$ & & & $0.0007$ & \\
 & $\textrm{E2}_{\textrm{norm}}$ & & & $0.01 \pm 0.1$ & & & $\approx 10^{-6}$ & \\
 & $\textrm{E2}_{\textrm{unif}}$ & & & $0.01 \pm 0.1$ & & & $\approx 10^{-6}$ & \\
 & $\textrm{A1}$ & & & $< 0.01$ & & & $ - $& \\ 
 & $\textrm{E1}$ & & & $< 0.01$ & & & $ - $ & \\ 
\enddata
\tablenotetext{}{\textbf{Notes.} The last column shows the percentage probed with respect to the total number of sources in-band for each model (second column of Tab.~\ref{tab:inband}). The remaining models, here indicated generally as A1 and E1, give no detectable signal in any of the realisations.}
\end{deluxetable}

Amongst all our models, only those where ellipticity is drawn as outlined by model 2 (Sec. \ref{sec:scen2}) give currently detectable objects, albeit just barely and with big uncertainties. Table \ref{tab:detect} summarises the results. We will be talking about agnostic models first.

\begin{figure}
\caption{
2D histogram showing $\overline{n}_{\textrm{bin}}(B,\varepsilon)$ color-coded for model $\textrm{A2}_{\textrm{high}}$.
In order to increase the resolution, we considered a higher number of realisations: 1000.
The white bins are bins without detected sources, hence $\overline{n}_{\textrm{bin}}<10^{-3}$. 
}
\centering
\includegraphics[width=\columnwidth]{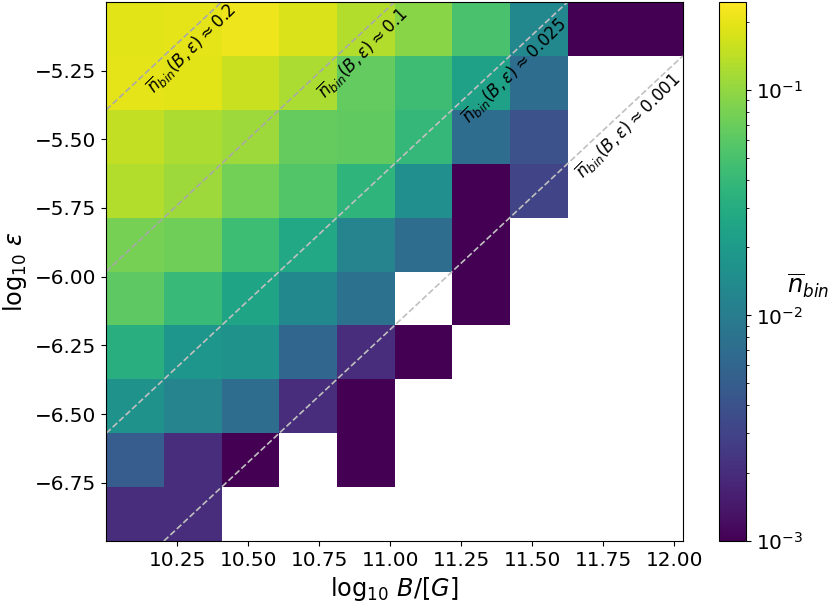}
\label{fig:demog_1}
\end{figure}

Considering model $\textrm{A2}_{\textrm{high}}$, which is the most populated by currently detectable sources, we look at the plot of Figure~\ref{fig:demog_1}. This shows how the average number $\overline{n}_{\textrm{bin}}(B, \varepsilon)$ of currently detectable sources depends on magnetic field $B$ and ellipticity $\varepsilon$; these two quantities are binned in log scale in 10 intervals in both dimensions.
When $\overline{n}_{\textrm{bin}}(B, \varepsilon) < 1$, it can be interpreted as the probability to find a detectable source with magnetic field and ellipticity within the ranges spanned by that bin.
We find that the constant $\overline{n}_{\textrm{bin}}$ curves (grey dashed lines in the plot) are well described by straight lines of the form
\begin{equation}
\label{eq:isoNumberDetectableSources}
\log_{10} \varepsilon = a \log_{10} B/[G] + b(\overline{n}_{\textrm{bin}})
\end{equation}
where $a\approx 1$ and $b$ depends on the iso-probability value $\overline{n}_{\textrm{bin}}(B, \varepsilon)$. 

We can invert Equation \ref{eq:isoNumberDetectableSources} and find a functional form for $\overline{n}_{\textrm{bin}}(B, \varepsilon)$. In order to do so we study how $b$ depends on $\overline{n}_{\textrm{bin}}$ and find $b(\overline{n}_{\textrm{bin}})$ to be well approximated by a quadratic polynomial, that is monotonic in the range of values shown in the colorbar of Figure~\ref{fig:demog_1}. We hence substitute the quadratic $b(\overline{n}_{\textrm{bin}})$ in Equation~\ref{eq:isoNumberDetectableSources}, solve for $\overline{n}_{\textrm{bin}}$ as a function of $B$ and $\varepsilon$ and find
\begin{equation}
\label{eq:n_of_B_eps}
\overline{n}_{\textrm{bin}}(B, \varepsilon) = 0.245 (1 - \sqrt{g(B,\varepsilon)} ),
\end{equation}
where
\begin{equation}
\label{eq:Delta_B_eps}
g(B, \varepsilon)  = 0.56 ~(a \log_{10} B/[G] - \log_{10} \varepsilon) - 8.39.
\end{equation}
We stress that Equation~\ref{eq:n_of_B_eps} was empirically derived based on the results shown in Figure~\ref{fig:demog_1}, so its validity outside of that $(B,\varepsilon)$ range has not been verified. Furthermore it is not valid in the ``white region" of Figure~\ref{fig:demog_1}, corresponding to $\overline{n}_{\textrm{bin}} < 0.001$, for which it yields negative values. On average in its range of validity the relative difference between our fit and the original results is 65\%.

We now turn to a qualitative explanation on why the constant detection-probability lines in the $(\varepsilon , B)$ are of the form $\varepsilon \propto B$.

From Equation \ref{eq:combsd}, setting $\sin \chi = 1$, we find that the time it takes a neutron star born with initial spin frequency $\nu_0$ to spin down to frequency $\nu$ as a function of its ellipticity and magnetic field, is 
\begin{equation}
\label{eq:combsdintegr}
\begin{split}
\tau(\nu; \nu_0, \varepsilon, B)
= & \frac{1}{2 |\gamma_{dip}|} \Bigg[ \frac{\nu_0^2 - \nu^2}{\nu_0^2 \nu^2} + 
\gamma \ln \left( \frac{\nu^2}{\nu_0^2} \left(  \frac{1 + \gamma\nu_0^2 }{1 + \gamma\nu^2 } \right) \right) \Bigg],
\end{split}
\end{equation}
where from Equations~\ref{eq:dotnu} and \ref{eq:emsdcond} we recall that $\gamma_{dip}\propto B^2$ and $\gamma\propto {\varepsilon^2\over {B^2}}$.

\begin{figure}
\caption{Spin-down time $\tau$ as from equation \ref{eq:combsdintegr} as a function of the magnetic field.}
\centering
\includegraphics[width=\columnwidth]{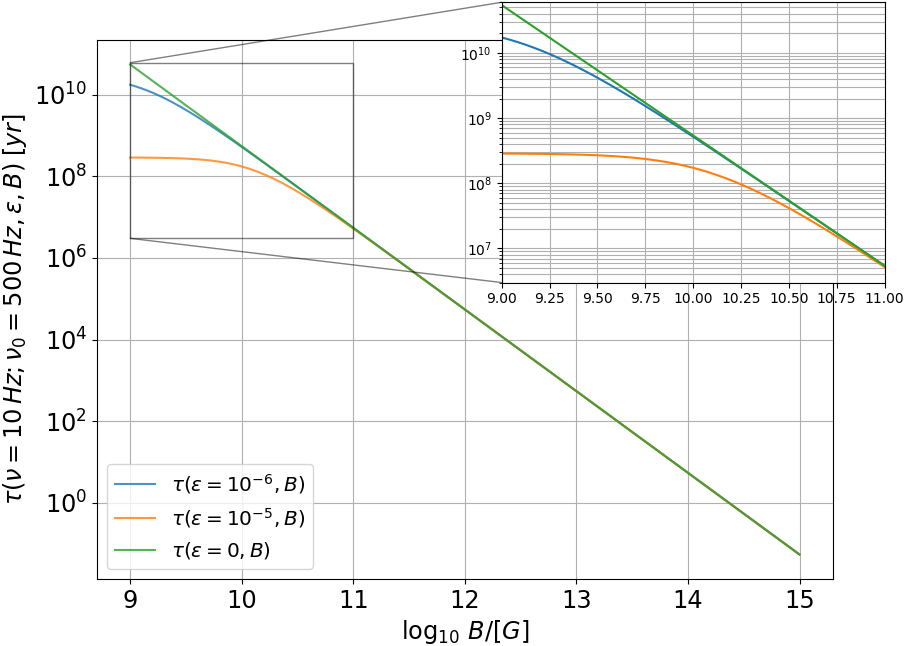}
\label{fig:spindowntime}
\end{figure}

Figure \ref{fig:spindowntime} shows $\tau(B)$ for $\nu=10$ \textrm{Hz}, $\nu_0=$ 500 \textrm{Hz} as a function of $B$ for various values of the ellipticity $\varepsilon$. 
For high values of $B$ the second term in the RHS of the equation above is negligible, and $\log{\tau}\propto -\log{B}$ (i.e. the plot is a straight line of slope $-1$). 
Only for $B\lesssim 10^{10}$ G and $\varepsilon \gtrsim 10^{-6} $ the loss of energy through gravitational waves due to the ellipticity becomes resolvable, with obviously longer evolutions for lower ellipticities. 

Our agnostic models all have $B \geq 10^{10}$ G, where the spin-down evolution is \emph{mostly} dominated by the magnetic field.
In this regime if $\nu_0\gg\nu$ then $\tau$ is largely independent of $\nu_0$ and we can take the curves of Figure \ref{fig:spindowntime} to be representative of the spin-down time for a signal to reach $10\,\textrm{\textrm{Hz}}$, starting from a sufficiently high birth spin frequency (indicatively above $100\,\textrm{Hz}$). This says that all sources older than
\begin{equation}
\tau^{age} \approx {1\over {2 |\gamma_{dip}|\nu_{min}^2}} = 5.4\times 10^{8} ~\textrm{yr} \left( {
{10^{10}~\textrm{G} } \over {B} 
} \right)^2 
\left({
{10~\textrm{\textrm{Hz}}}\over {\nu_{min}}
}\right)^2
\label{eq:agemaxfreq}
\end{equation}
are rotating slower than $\nu_{min}$.

How many stars are there younger than $\tau^{age}$?
Since we assume a constant birth rate of $\mathcal{R} \approx 4 \times 10^{-2}$ yr$^{-1}$, the number of stars younger than $\tau^{age}$ is simply the number of objects born between now and a time $\tau^{age}$ back in the past:
\begin{equation}
\label{eq:NolderAge}
N(\tau^{age}) = \mathcal{R} \cdot \tau^{age} \approx 2.16 \times 10^{7} \left( { {10^{10}~\textrm{G} } \over {B} } \right)^2 \left({ {10~\textrm{\textrm{Hz}}}\over {\nu_{min}} }\right)^2.
\end{equation}
This is also approximately the number of objects spinning {\it{faster}} than $\nu_{min}$.

From Equation~\ref{eq:NolderAge} we see that if $B$ increases, the number of objects spinning faster than $\nu_{min}$ decreases $\propto B^{-2}$. On the other hand, the total number of objects within a given distance $N(d)$ is approximately proportional to $d^2$ (this proportionality is exact if the neutron stars are distributed on a plane), and out of these only the ones above some $h_0$ value are going to be detectable. So with increasing $B$ the overall number of objects in band decreases, but the number of detectable ones could be kept constant by increasing $\varepsilon\propto d$, from Equation~\ref{eq:h0}. This is the reason why the lines of constant number of detectable sources have slope $\approx 1$ in the $\log B-\log\varepsilon$ plane, with the low $B$/high $\varepsilon$ combination, being the most favourable for detection, as also found by \cite{Wade2012}.
 
\subsubsection{Impact of birth spin frequency} \label{sec:birth_spin_freq}

Comparing results from models $\textrm{A2}_{\textrm{low}}$ and $\textrm{A2}_{\textrm{high}}$ tells us something about the impact of birth spin frequency on detectability. The difference between the two models is that the highest birth spin frequency in model ``low" is 300 Hz whereas in model ``high" is 1200 Hz. 
Since birth spin frequencies are distributed log-uniformly, only $\approx 22\%$ of neutron stars in $\textrm{A2}_{\textrm{high}}$ have $\nu_0 \in [300\,\textrm{Hz}, 1200\,\textrm{Hz}]$, yet these account for $\approx 66\%$ of the detectable sources. This means that, under the astrophysical assumptions outlined by our agnostic (A) models, the probability of a neutron star born with spin frequency above $300\,\textrm{Hz}$ to emit a currently detectable signal, is about $\sim 7$ times larger than that of a neutron star born with spin frequency below $300\,\textrm{Hz}$. We explain the last statement explicitly in Appendix~\ref{sec:factor_7}. \subsubsection{Phase parameters and sky distribution} \label{sec:phase_param}

\begin{figure}
\caption{$f_{GW} - \dot{f}_{GW}$ distribution of strong signals, that is, signals that are currently detectable (filled opaque stars) or within a factor of 3 of being detectable (empty transparent stars).
The coloured areas correspond to the parameter space portions covered by all-sky searches mentioned in Sec.~\ref{sec:detect}. For the few sources that fall outside of the $f-\dot{f}$ region covered, the amplitude is compared to the upper limit set by the \emph{FrequencyHough} pipeline in \cite{2022PhRvD.106j2008A} at the frequency closest to the signal frequency. The $(f_{GW},\dot{f}_{GW})$ of putative sources evolving due to purely electromagnetic emission are shown by the dotted lines and due to gravitational braking by the dash-dotted lines. The strong sources in the white areas -- i.e. not covered by any search --  amount to less than 0.5\% of all sources in the plot. E2 signals are marked with a bigger and thicker star. The hatched area is ignored in models A since we set $B_{min} = 10^{10}\,\textrm{G}$ in these models.}
\centering
\includegraphics[width=\columnwidth]{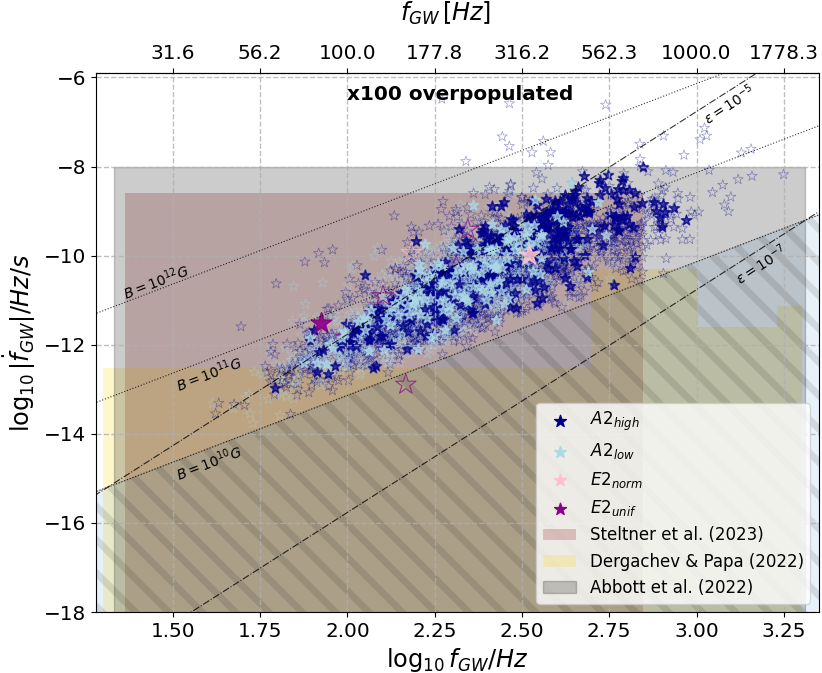}
\label{fig:ffdot_contour}
\end{figure}

Figure~\ref{fig:ffdot_contour} shows how the phase parameter space region covered by the different searches compares to the region occupied by the loudest signals (the definition of ``loud" signal is given in the caption of the figure). The latter is overall well covered by ongoing efforts.
There is a very scarcely populated region in this plot, that is routinely searched but that in principle could be dropped, based on our results: the high $\dot{f}$ region, at frequencies $\lesssim 130$ \textrm{Hz}.  Since however the computational cost scales at least with the square of the frequency, the overall benefit gained by excluding the region at issue would not result in a significant saving in resources. It is hence unlikely that the savings coming from becoming completely blind to signals from this region could be fruitfully re-invested to significantly increase the sensitivity of the searches in other regions of the parameter space.

\begin{deluxetable}{ccc}
\label{tab:bands}
\tablecaption{Average number of detectable objects per frequency band for model $\textrm{A2}_{\textrm{high}}$.}
\tablehead{ 
				\emph{low}         &              \emph{mid}         &         \emph{high}			      \\
	               $[20, 100]~\textrm{Hz}$ & $[100, 500]~\textrm{Hz}$  & $[500, 2400]~\textrm{Hz}$  } 
\startdata
                 \multicolumn{1}{c}{}  &\multicolumn{1}{c}{\textbf{now}}&\multicolumn{1}{c}{}  \\
                               0.12           &        3.08               &          0.42            \\ \hline \hline
                  \multicolumn{1}{c}{}  &\multicolumn{1}{c}{\textbf{$2 \times$ more sensitive}}&\multicolumn{1}{c}{}\\
                                  0.69          &        11.26            &          1.37            \\ \hline \hline
                  \multicolumn{1}{c}{}  &\multicolumn{1}{c}{\textbf{$3 \times$ more sensitive}}&\multicolumn{1}{c}{}\\
                                   1.79         &        23.23             &          2.5            \\ \hline \hline
                  \multicolumn{1}{c}{}  &\multicolumn{1}{c}{\textbf{$10 \times$ more sensitive}}&\multicolumn{1}{c}{}\\
                                    33.77       &        170.12            &         9.74           \\ 
\enddata
\end{deluxetable}

Table \ref{tab:bands} shows the expected number of detectable objects in various frequency ranges, having taken the $\textrm{A2}_{\textrm{high}}$ as our reference model, as done before. The table also shows how the expected number of detectable signals changes as the detectors' sensitivity increases. 
There are a number of competing factors that determine the detectability in different bands. Namely, the frequency-dependence on the detector noise, the likelihood of having a source in any band, and the proportionality between the amplitude of a signal and the square of its frequency.
The detector sensitivity curve favours the \emph{mid} band, signal occupancy favours the \emph{low} band and the frequency dependence of signal amplitude favours the \emph{high} band. 

\begin{figure}
\caption{Distribution of the frequency of signals from model $\textrm{A2}_{\textrm{high}}$ that are detectable now and with higher (tenfold) sensitivity detectors/searches. The vertical solid lines located at 162 and 287 Hz indicate the median of the same-colour histogram.}
\centering
\includegraphics[width=\columnwidth]{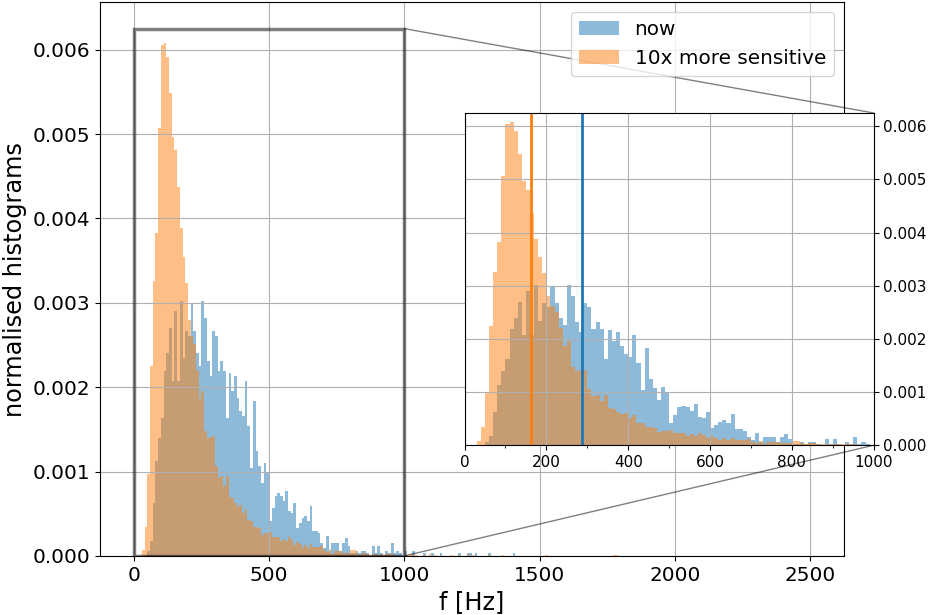}
\label{fig:freq_hists}
\end{figure}

Our results suggest that the \emph{mid} frequency band lays on a ``sweet spot" and is the major contributor to the overall detectability. Also, at current sensitivity, the \emph{high} band contributes significantly more than the \emph{low} band. However, as both Table \ref{tab:bands} and figure Figure \ref{fig:freq_hists} show, this trend slowly reverses as the detectors' sensitivity increases, and is definitely inverted at the sensitivities 10 times higher than the current ones, i.e. the level expected for next generation of ground-based detectors. This effect is obviously amplified for detectors -- like the Einstein Telescope -- that promise significantly improved performance at low frequencies.

The smallest band that currently contains $90\%$ of the detectable signals is the band $[65, 545]\,\textrm{Hz}$, with this interval shrinking and shifting to the left to $[45, 380]\,\textrm{Hz}$ in the case of detectors/searches $10\times$ more sensitive than current ones.

\begin{figure}
\caption{Distribution on the sky (Galactic coordinates) of signals that are detectable (filled opaque stars) and within a factor of 3 of being detectable (empty transparent stars), as in Figure \ref{fig:ffdot_contour}.}
\centering
\includegraphics[width=\columnwidth]{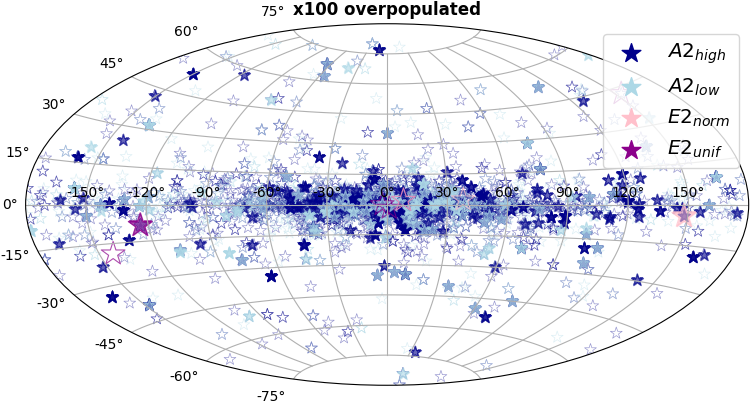}
\label{fig:skymap}
\end{figure}

All \emph{loud} signals shown in Figure \ref{fig:ffdot_contour}, are also plotted in the sky-map of Figure \ref{fig:skymap}. 
We consider the region within $15^{\circ}$ the galactic plane. It contains $\approx 81\%$ of all loud signals of $\textrm{A2}_{\textrm{low}}$ model and $\approx 86\%$ of all loud signals of $\textrm{A2}_{\textrm{high}}$ model (we ignore signals from $\textrm{E2}_{\textrm{norm}}$ and $\textrm{E2}_{\textrm{unif}}$ because of their small statistical sample size).
These percentages are significantly higher than the percentage of objects within the same region from the entire synthetic neutron star population, that is $63.5\%$. This is due to the fact that loud signals come by and large from young sources, and since most neutron stars are born within a narrow region of the galactic plane, the young ones have not had enough time to migrate away from it. 

Coming back to the fact that, depending on the model, $\approx 81 - 86\%$ of all loud signals lay within $15^{\circ}$ the galactic plane, a similar percentage (about $85 \%$) is found within the same sky region in the ATNF catalogue \citep{Manchester2005} if pulsars in globular clusters as well as recycled ones are discarded. Since -- similarly to the gravitational wave case -- also EM detections from isolated non-recycled pulsars are subject to selection bias towards young objects, this last fact can be considered a consistency check of our synthetic population.

\subsubsection{Models E2}
\label{sec:models_E2}
To conclude this section we briefly comment on results from models E2. 

For both $\textrm{E2}_{\textrm{norm}}$ and $\textrm{E2}_{\textrm{unif}}$ only one signal out of the 100 realisations was found to be detectable. This might appear surprising given that the E-model populations have more stars in-band than the A-model ones, see Table~\ref{tab:inband}. The reason for such can be explained as follows. Assuming that the contribution of the ellipticity to the spin-evolution is negligible and assuming a constant magnetic field, Equation~\ref{eq:agemaxfreq} says that the time it takes a system to spin down to frequency $\nu$ from a much higher starting frequency is proportional to $1/B^2$. Hence, if the magnetic field is not constant but decays, it takes a longer time for the star to spin down to $\nu$. 
This happens if the spin evolution time-scale (based on the initial magnetic field) is longer than the magnetic field decay time $\tau_B$, since in this case the magnetic field decays appreciably during the spin-evolution. So when 
\begin{equation}
\begin{split}
\label{eq:Espindowntime}
5.4\times 10^{8} ~\textrm{yr} \left( {{10^{10}~\textrm{G} } \over {B_0} }\right)^2   \left( {{10 {\textrm{Hz}}\over{\nu_{min}} }}\right)^2
> \tau_B
 \Rightarrow \\
 \Rightarrow B_0 \leq 2.3\times 10^{11}~\textrm{G} ~\left[ {{10 ~{\textrm{Hz}} }\over{{\nu_{min}}}}\right]
 \end{split}
\end{equation}
the star will stay in band indefinitely\footnote{In fact for the E-models we have $\approx 10^7$ objects with $B_0 < 2.3 \times 10^{11}$ G, which is consistent with the $\approx 10^6$ in-band objects from Table \ref{tab:inband}, considering that not all spins at birth are very high.}. A star like this will however not be detectable because ellipticities generally higher than $10^{-6}$ are necessary for detection today, and objects with such large ellipticities would have spun-down to frequencies below $10 \, \textrm{Hz}$ -- see zoomed-inset of Figure~\ref{fig:spindowntime}, remembering that, under the assumption of constant birthrate, 99\% of neutron stars today are at least $10^8$ years old.

However, as both searches and detectors' sensitivities improve, the chances to be able to detect a less deformed object increase.
Indeed the loudest signals of E2 models are not too far below current upper limits, and, as it will be discussed in the next section, these models become interesting for third generation detectors (see also the summary-figure, Figure~\ref{fig:2dhists_scen2}, at the end of the paper).

\subsection{Third generation detectors}\label{sec:thirdgendet}

We now want to investigate on the prospects of detection by the third generation ground-based gravitational wave detectors Einstein Telescope (ET) \citep{Punturo2010, Maggiore2020} and Cosmic Explorer (CE) \citep{Dwyer2015}. 
For ET we use the sensitivity estimations presented in \cite{Hild2011} there labelled as ``ET-D" in the equilateral triangle configuration, while for CE we use the single detector 40 km ``baseline" configuration as defined in \cite{Srivastava2022}. 

In order to assess the detectability of a population of signals using the estimated sensitivity of proposed detectors we proceed as follows. As a measure of the sensitivity of a search we take the upper limit values, and from those we define the sensitivity depth \citep{Behnke2015}: 
\begin{equation}
\label{eq:sensDepth}
\mathcal{D}^{C} \equiv \frac{\sqrt{S_h(f)}}{h_0^{C}(f)}~~[1/\sqrt{{\textrm{\textrm{Hz}}}}]
\end{equation}
where $h_0^{C}(f)$ are the amplitude upper limits at confidence level $C$ and $S_h(f)$ is the noise spectral density. 
The sensitivity depth is a property of the search and it measures how deep a certain search method could ``dig" into given detector noise.

Assuming to perform the same search on ET and CE data as was performed on LIGO data, we estimate the expected upper limits simply by solving Equation~\ref{eq:sensDepth} for $h_0^{95\%}$ by using the predicted 3rd-generation detectors $S_h(f)$ and the sensitivity depth ${\mathcal{D}^{95\%}}$ of the LIGO-data search: 
\begin{equation}
\label{eq:rescaling}
h_0(f)^{95\%} = \frac{\sqrt{S_h(f)}}{\mathcal{D}^{95\%}}\sqrt{{2\over {N}}}.
\end{equation}
$N$ is the number of 3-rd generation detectors equivalent to the number of detectors used in the LIGO-data search. The three-arm design of ET is equivalent to a system of three Advanced generation detectors \citep{Hild2011}, so $N^{ET}=3$, whereas we conservatively take $N^{CE}=1$\footnote{The main 40 km CE observatory might be combined with a 20 km detector. In such case however we cannot simply consider $N^{CE} = 2$ since Equation \ref{eq:rescaling} is valid only when the various detectors are of comparable sensitivity.}. 

We can now compare the predicted upper limits from Equation~\ref{eq:rescaling} with our synthetic signal population, and see how many are detectable. We simplify the detectability criteria such that any signal whose amplitude is bigger than or equal to the upper limit at the frequency of the signal is considered to be detectable, regardless of the rest of the signal parameters. Of the signals that result to be detectable according to this criteria, $\approx 99\%$ have $|\dot{f}|<10^{-8}\,\textrm{Hz}/s$ (so within the current searched range), so this simplification does not introduce any significant bias.

\begin{deluxetable}{lcccccc}
\label{tab:detect_thirdgen}
\tablecaption{Average number of sources detectable by ET and CE with searches comparable with the advanced-LIGO data searches \citep{2022PhRvD.106j2008A,2023ApJ...952...55S}.}
\tablehead{\colhead{Model} & & & \colhead{$\overline{n}$} & & & $\%$ of \emph{in-band}\\
 & & \colhead{ET} & & \colhead{CE} & } 
\startdata
 $\textrm{A2}_{\textrm{low}}$ & & $231.9 \pm 14.6$ & & $338.1 \pm 16.8$ & & $0.003\%$\\
 $\textrm{A2}_{\textrm{high}}$ & & $387.2 \pm 19.4$ & & $524.3 \pm 22.6$ & & $0.005\%$\\
 $\textrm{E2}_{\textrm{norm}}$ & & $0.5 \pm 0.6$ & & $2.0 \pm 1.4$ & & $0.00001\%$\\
 $\textrm{E2}_{\textrm{unif}}$ & & $1.7 \pm 1.3$ & & $5.2 \pm 2.2$ & & $0.00002\%$\\
\enddata
\tablenotetext{}{\textbf{Notes.} The last column is obtained considering the total detectable by either ET or CE and, similarly to Table~\ref{tab:detect}, represents the fraction of detectable sources over the total number of sources in-band for each model (third column of Table \ref{tab:inband}).}
\end{deluxetable}

\begin{figure*}
\caption{Normal neutron stars: 2D histograms showing the expected number of sources at each $f_{GW}-h_0$ bin, $\overline{n}$. There are 50 log-uniform bins in frequency and $h_0$. The frequency interval is $[2\,\textrm{Hz} - 2100\,\textrm{Hz}]$; the $h_0$ interval is $[10^{-34} - 10^{-27}]$. The curve for current detectors is based on published searches for isolated neutron stars. The ``3gen detectors" line is a sensitivity projection based on forecast noise curves and assumes search methods as sensitive as current searches. At each frequency, we use the best sensitivity between the ET and the CE detectors, as explained in Section \ref{sec:thirdgendet}. The“50x better 3gen detectors” line is the“3gen detectors” line divided it by 50.}
\centering
\includegraphics[width=\textwidth,height=\textheight,keepaspectratio]{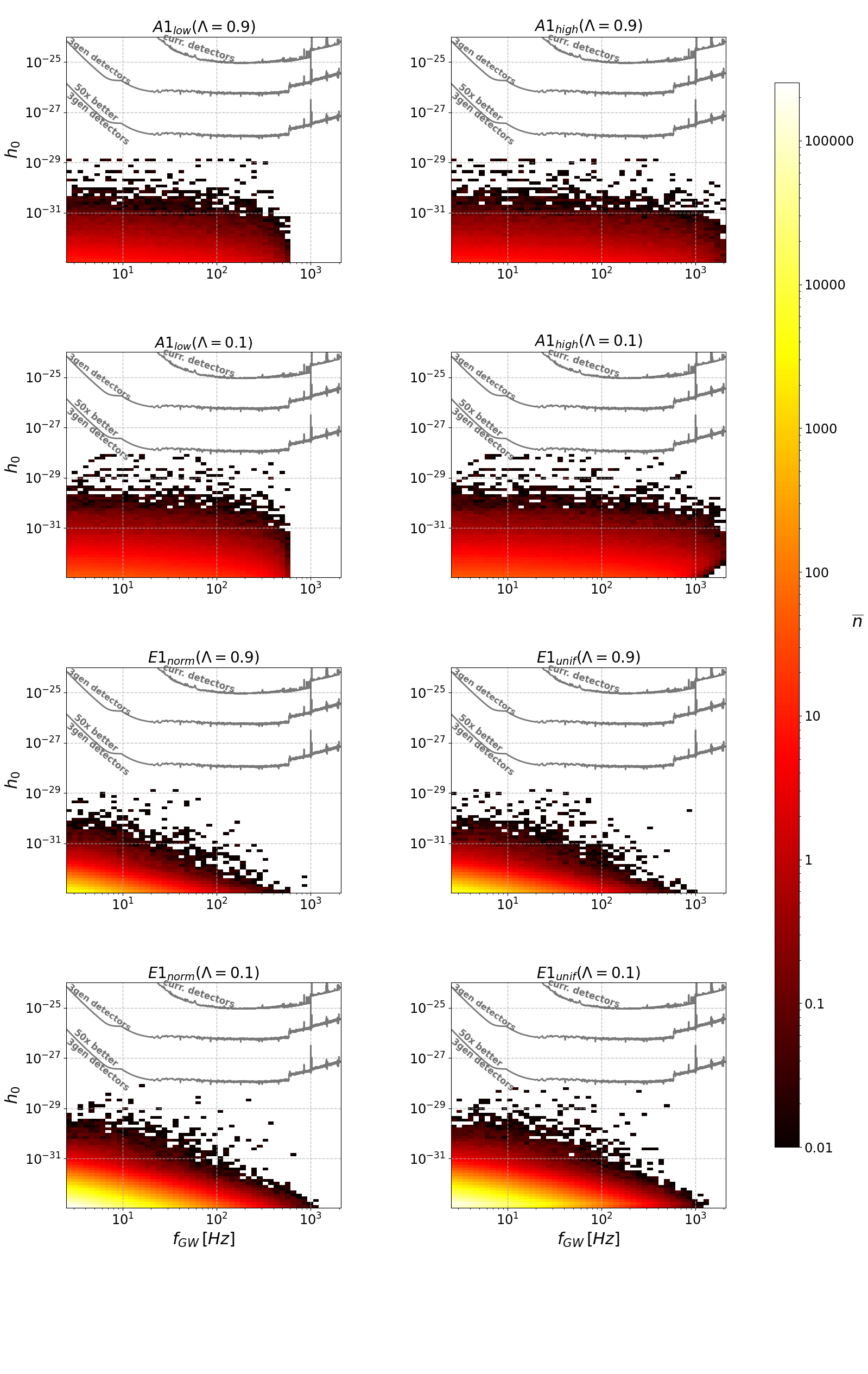}
\label{fig:2dhists_scen1}
\end{figure*}

\begin{figure*}
\caption{Same as Figure \ref{fig:2dhists_scen1}, but for models for which the ellipticity is drawn from a log-uniform distribution, as explained in Sec. \ref{sec:ellip} for Model 2. The frequency interval is $[2\,\textrm{Hz} - 2100\,\textrm{Hz}]$; the $h_0$ interval is $[10^{-28} - 10^{-23}]$.}
\centering
\includegraphics[width=\textwidth]{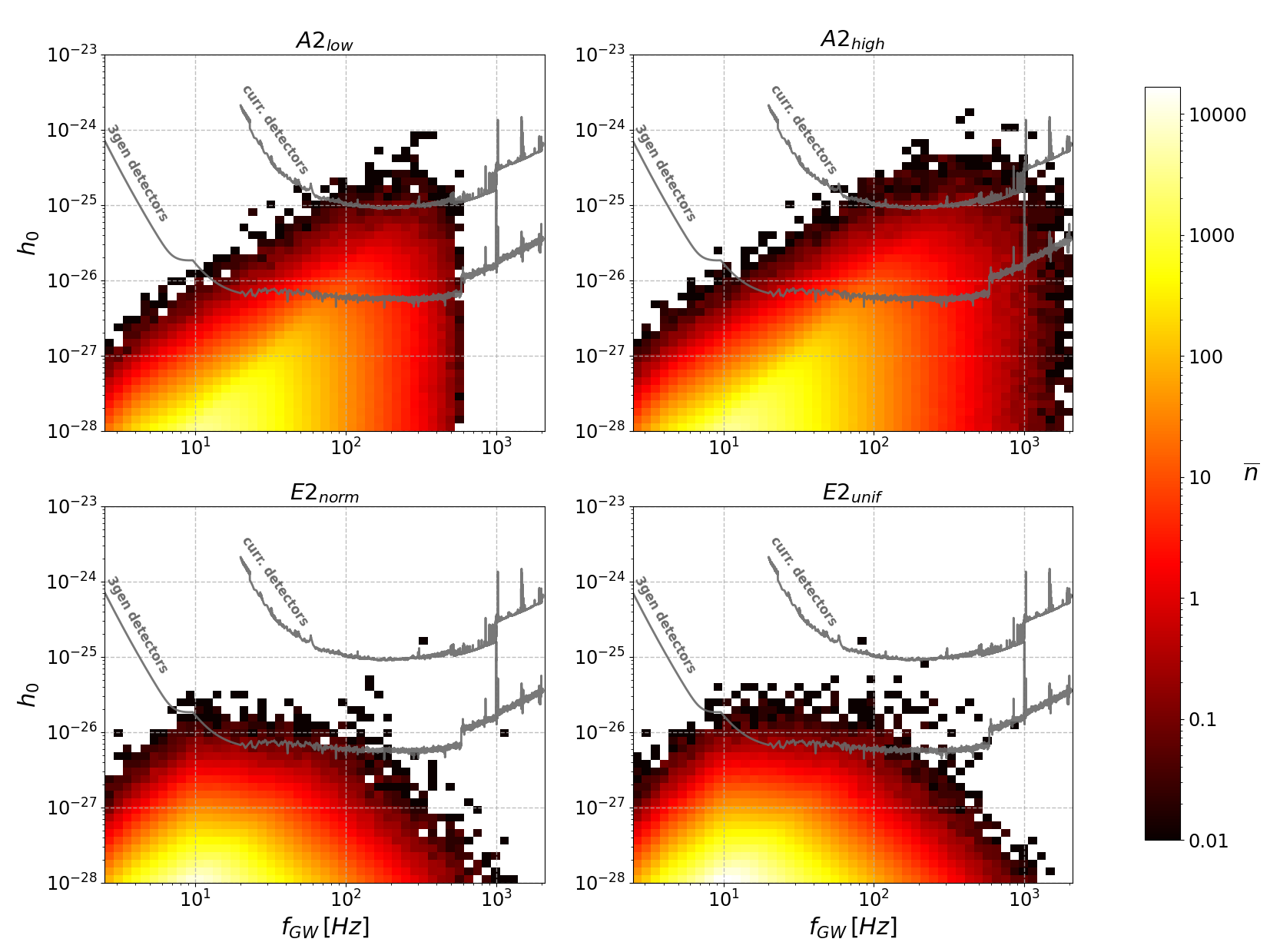}
\label{fig:2dhists_scen2}
\end{figure*}

Results are summarised in  Table \ref{tab:detect_thirdgen}. For a more detailed picture we refer the interested reader to the summary Figures \ref{fig:2dhists_scen2} and \ref{fig:2dhists_scen1}, at the end of the paper.
If the ellipticity is purely generated by the magnetic field as per the A1 and E1 models, not even the 3rd-generation gravitational wave detectors are likely to see a signal. Conversely, all of our models where ellipticity is log-uniformly distributed up to a maximum value of $10^{-5}$ (Model 2 defined in \ref{sec:ellip}) give detectable signals. 

\begin{deluxetable}{clc|cccccc}
\label{tab:bands_3gen}
\tablecaption{Average number of detectable objects per frequency band in different models for CE.}
\tablehead{\multicolumn{1}{c}{}&\multicolumn{1}{c}{Model}     &\multicolumn{1}{c}{}&&\multicolumn{1}{c}{\emph{low}} &&\multicolumn{1}{c}{\emph{mid}}&&\multicolumn{1}{c}{\emph{high}} \\
		\multicolumn{1}{c}{}&\multicolumn{1}{c}{}     &\multicolumn{1}{c}{}&&\multicolumn{1}{c}{$[5,100]\textrm{Hz}$} &&\multicolumn{1}{c}{$[100,500]\textrm{Hz}$}&&\multicolumn{1}{c}{$[500,2500]\textrm{Hz}$}} 
\startdata
                                                 &$\textrm{A2}_{\textrm{low}}$&				     &&                          151.59                        &&                           186.73                        &&                              0.53                         \\
                                                 &$\textrm{A2}_{\textrm{high}}$&				     &&                          193.2                        &&                           318.02                        &&                              15.15                         \\
                                                 &$\textrm{E2}_{\textrm{norm}}$&		             &&                           1.83                        &&                           0.21                        &&                              0.0                         \\
                                                 &$\textrm{E2}_{\textrm{unif}}$&				     &&                          4.38                        &&                           0.78                        &&                              0.02                         \\                                                                                                                                                   
\enddata
\end{deluxetable}

Table \ref{tab:bands_3gen} shows the average number of sources detectable by 3rd-generation detectors in three different frequency ranges, and highlights the importance of better low-frequency sensitivity. Compared to the ranges defined in Sec. \ref{sec:phase_param}, the low frequency band is now pushed down to $5$ Hz and the high band is pushed up to $2500$ Hz. The chances of a detection of a signal from population model $\textrm{A2}_{\textrm{high}}$ in the low band grows by a factor $\approx$ 1000-fold with respect to current detectors. 

\begin{figure} 
\caption{The gravitational wave amplitude of signals (stars) and the expected 3rd-generation detector sensitivity.
Magenta stars represent signals from model $\textrm{E2}_{\textrm{unif}}$, which includes magnetic field decay, but for this comparison, we show the signals from the same model, but with a static magnetic field (dark green stars).}
\includegraphics[width=\columnwidth]{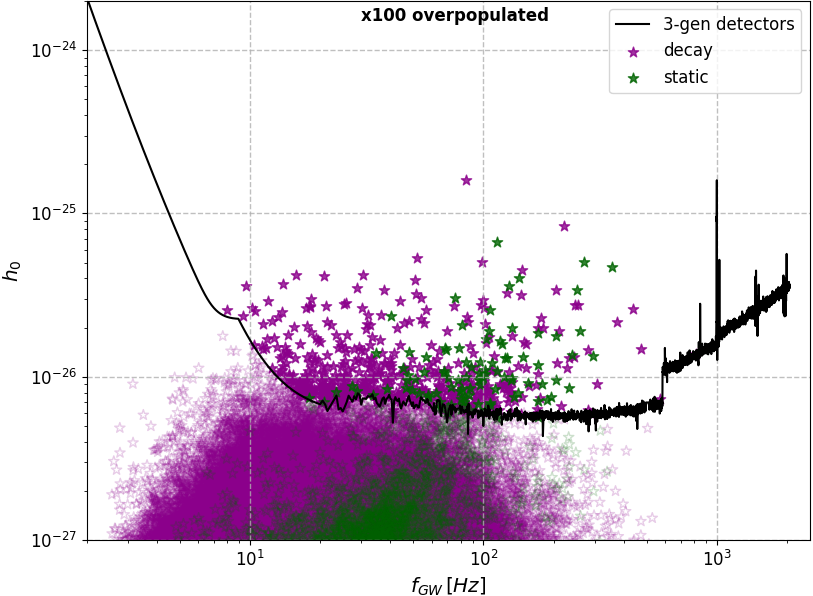}
\label{fig:decay_test}
\end{figure}
\begin{figure} 
\caption{Kernel density estimation plot of the number density of third-generation detectable sources as a function of magnetic field and age for the two populations shown in Figure \ref{fig:decay_test}. The dashed empty contours (in black and white) represent the magenta population objects at birth. The contour levels go from 0 to 100\% in steps of 10\%.}
\includegraphics[width=\columnwidth]{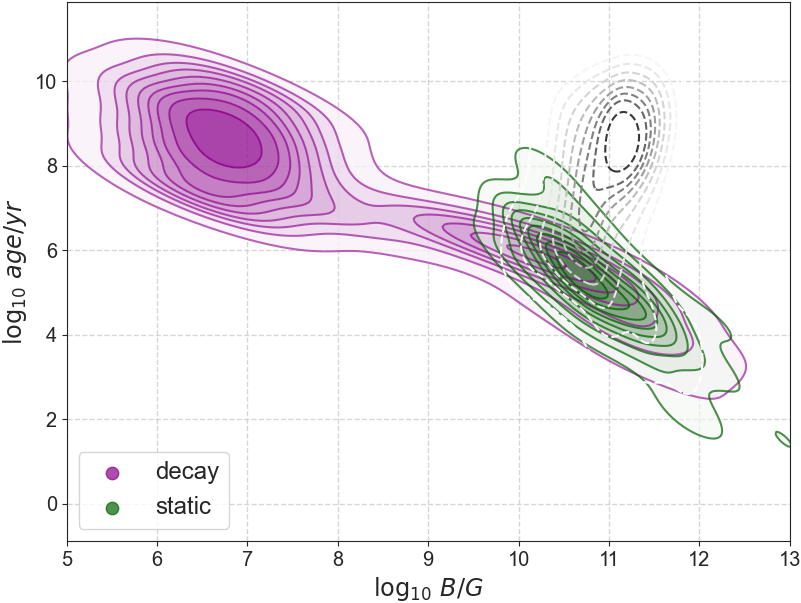}
\label{fig:decay_test1}
\end{figure}
The magnetic field decay introduced in model E2 has a large impact on the number of sources detectable by third generation detectors, enhancing the chances of detection as shown in Figure~\ref{fig:decay_test}. The magnetic field decay makes it possible for old sources (up to several billion years), born with fields $B \lesssim 10^{11.5} \, \textrm{G}$, to still be in-band (Equation~\ref{eq:Espindowntime}). This creates an {\it{additional}} population of detectable signals with respect to the static magnetic field population, which appears as the high-age concentration ``blob" in Figure~\ref{fig:decay_test1}. 
However, as already explained in Sec.~\ref{sec:models_E2}, such very old in-band sources cannot be maximally deformed ($\varepsilon \lesssim 3 \times 10^{-6}$).

\section{Recycled Neutron Stars}\label{sec:recycled}

In the past sections, we have considered the frequency evolution of populations of normal neutron stars from birth until now, and obtained synthetic present-day populations. We have seen that a major factor impacting the detectability of stars in these populations is their frequency, i.e. whether they have spun during their evolution down to frequencies too low to be detectable.

From electro-magnetic observations of pulsars we know that there exists another category of neutron stars -- so called ``recycled" objects --  rotating faster than the typical normal neutron star, which makes them very interesting for continuous gravitational wave detection. 

A recycled neutron star is an old neutron star that has been spun up to spin periods of the order of the milliseconds as a result of total angular momentum conservation during an accretion phase \citep{Alpar1982, RADHAKRISHNAN1982, Bhattacharya1991}. 
An example of such objects are millisecond pulsars (MSPs), that are visible in the electromagnetic spectrum as radio, X-ray and gamma sources. For a review on MSPs see \cite{Lorimer2008}.

Like for the spatial distribution of MSPs, the distribution of galactic recycled neutron stars might be quite different than that of normal ones. Moreover, the evolutionary path that links a recycled neutron star with its progenitor binary star system is complex, with the outcome of the recycling generally coupled with the binary parameters. Modelling such mechanism is beyond the scope of this paper. But since this population is so relevant for continuous waves, here we consider a simplified population of non-accreting fully recycled neutron stars \citep{tauris2011}, and use it to make a first detectability assessment and compare with the results for the population of normal neutron stars. 

For the spatial distribution, we opt for a simple ``snapshot" approach. Assuming the galactic population of recycled neutron stars to follow the MSPs one, we consider the spatial distribution that \citeauthor{Gregoire2013} adopt in their work, which is in turn based on the results of \citeauthor{Story2007}: 
\begin{equation}\label{eq:spatdistr}
p(\rho, z) \propto \exp{(-\rho/\rho_0)} \exp{(-|z|/z_0)},
\end{equation}
where $\rho_0 = 4.2 \, kpc$ and $z_0 = 0.5 \, kpc$ are the radial and vertical scale heights respectively. In \ref{eq:spatdistr} we use cylindrical coordinates with origin coincident with the centre of the galaxy; $\rho$ is the radial coordinate while $z$ is the axial coordinate. Azimuthal isotropy is assumed so the azimuth coordinate is uniformly distributed within $0$ and $2\pi$.

\begin{figure*}
\caption{Recycled systems: 2D histograms showing the expected number of sources at each $f_{GW}-h_0$ bin, $\overline{n}$. There are 100 log-uniform bins in frequency and $h_0$. The frequency interval is $[2\,\textrm{Hz} - 2100\,\textrm{Hz}]$; the $h_0$ interval is $[5\times 10^{-29} - 5\times 10^{-24}]$. The solid curves for current detectors refer to published searches for isolated neutron stars, and the dotted curves refer to published searches for neutron stars in binary systems. 
The ``3gen detectors" lines are sensitivity projections based on the forecast noise curves and assume search methods as sensitive as the current ones. At each frequency, we use the best sensitivity between ET and CE, as explained in Section \ref{sec:thirdgendet}.}
\centering
\includegraphics[width=\textwidth]{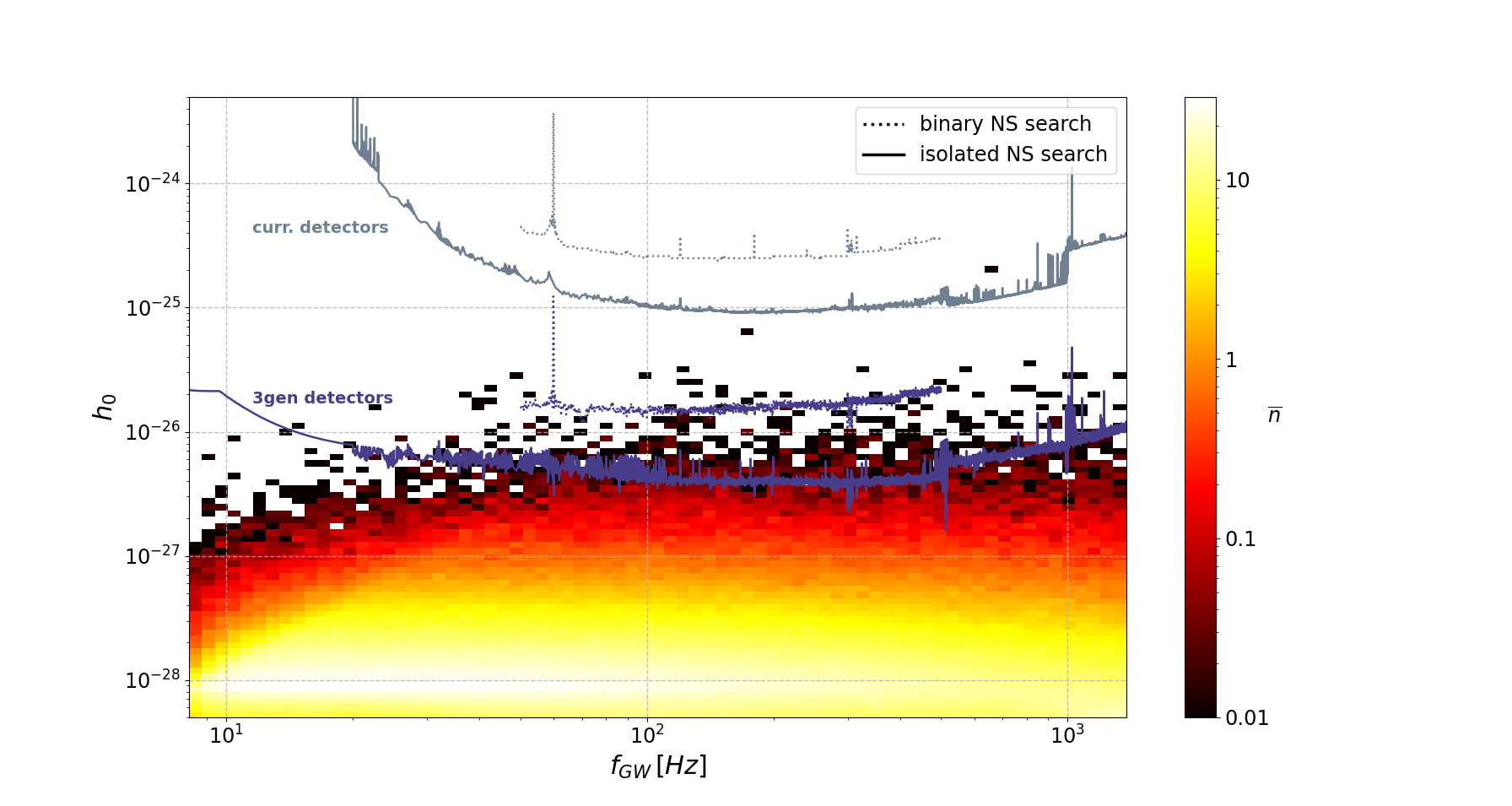}
\label{fig:2dhist_recycled}
\end{figure*}

We assume a constant birthrate of $5 \cdot 10^{-6}$ yr$^{-1}$ in the last $12$ Gyr\footnote{By ``birth" here we mean the birth of a recycled neutron star, i.e. immediately after the end of the recycling process.} resulting in a population of $N^{MSP} = 60\,000$ objects. The birthrate is obtained by \citeauthor{Story2007} from the total number of MSPs in the disk consistent with the detected population and is in loose agreement with those obtained by \citeauthor{Ferrario2007} and \citeauthor{Gregoire2013} (extended sample case). 

Magnetic fields are drawn from a log-uniform distribution between $10^7$ G and $10^9$ G \citep{Manchester2005}. 

For the ellipticity, we adopt the same models as described in Section \ref{sec:ellip}, but since external magnetic field values in recycled systems are not high enough to give a significant deformation through Equation~\ref{eq:mastrano}, we consider only ``Model 2" of Equation~ \ref{sec:ellip}. 

We let every object be recycled to the same initial spin frequency of $700 \, \textrm{Hz}$ that roughly corresponds with the observed cut-off frequency value in accreting millisecond X-ray pulsars \citep{Chakrabarty2008}, and evolve it through Equation~\ref{eq:combsd}. We find a much higher fraction of objects in band, compared to normal neutron stars: $N_{f > 20\textrm{Hz}}/N^{MSP}=0.96$ and $N_{f > 5\textrm{Hz}}/N^{MSP}=1$ for recycled objects versus the less than 1\% and 10\%, for the 20Hz and 5Hz cut-off respectively, for normal neutron stars (See Table~\ref{tab:inband}). This is consistent with the much lower magnetic fields assumed for recycled neutron stars. We say rightaway that these higher fractions of in-band objects will not translate in a higher detection probability for recycled systems, due to the overall much smaller number of objects.

Fully recycled neutron stars are observed both in binary systems and isolated. For this reason, to establish detectability, in addition to the searches for signals from isolated objects of Sec.~\ref{sec:detect}, we also consider the following upper limits from all-sky searches for neutron stars in binaries: 
\begin{itemize}
\item \cite{Covas2020} based on Advanced LIGO O2 data
\item \cite{Abbott_binary_2021} based on Advanced LIGO O3a data
\item \cite{Covas:2022rfg} based on Advanced LIGO O3a data.
\end{itemize}
The per-Hz cost of these searches is considerably higher than for all-sky searches for signals from isolated objects, because the parameter space includes at least the three binary parameters: orbital period, projected semi-major axis and time of ascending nodes. Since one operates at a limited computing budget, this results in a lower sensitivity. We have not considered \citep{Singh:2022hfd}, which presents the most stringent upper limits for continuous waves from binary systems, but with orbital parameters unlikely to pertain to recycled neutron stars. 

We determine the detectability of isolated neutron stars and neutron stars in binaries, separately. We count as detectable any isolated signal whose amplitude lays above the most sensitive upper limit set at the signal's frequency by the isolated searches listed in Section~\ref{sec:detect}, independently of the $\dot{f}$ range covered by that search. We count as detectable any binary signal whose amplitude lays above the most sensitive upper limit set at the signal's frequency by any of the binary searches listed above, independently of the $\dot{f}$ and orbital parameter ranges covered by that search. This is a reasonable assumption, because the most sensitive isolated and binary searches, each in its own category, are very close in sensitivity to each other, independently of their target $\dot{f}$ and orbital parameter ranges.

The sensitivity estimation of 3rd-generation detectors is done by rescaling the search sensitivity depth, as explained in the previous section. But differently than what done in the case of normal neutron stars, here for isolated objects we also consider the Falcon search, since the $\dot{f}$ range surveyed by this search is compatible with recycled neutron stars.

We also consider a mixed population with 58\% neutron stars in binaries and 42\% isolated, based on the fraction of binary-to-isolated MSPs. The results are shown in Figure \ref{fig:2dhist_recycled} and summarized in Table \ref{tab:msps}.

\begingroup
\setlength{\tabcolsep}{10pt}
\begin{deluxetable}{c|ccc}
\label{tab:msps}
\tablecaption{Average number of detectable sources $\overline{n}$ under different assumptions on the proportions of the number of binary-to-isolated neutron stars.}
\tablehead{ & & \colhead{$\overline{n}$} & \\
& \colhead{(all binary)} & \colhead{(all isolated)} & \colhead{(mix)} }
\startdata
 now & $< 0.01$ & $0.01\pm 0.1$ & $<0.01$\\
 CE & $0.18\pm 0.38$ & $5.8\pm 2.62$ & 3.44 \\
 ET & $0.09\pm 0.29$ & $4.76\pm 2.32$ & 2.8  \\
\enddata
\tablenotetext{}{\textbf{Notes.} In the mix-model we have assumed 42\% isolated objects and 58\% in binaries, consistent with the ATNF MSP population. We recall that the total population is 60,000 objects.}
\end{deluxetable}
\endgroup

We find no currently detectable signals from neutron stars in binaries in 100 realisations of the population, and only one out of 100 realisations assuming all the population is composed of isolated neutron stars. 
With 3rd-generation detectors, assuming a mixed population of isolated/binary recycled neutron stars, the average number of detectable signals varies approximately between $0.2$ and $6$, with the exact value depending on the relative fraction of the two populations. Assuming the same relative fraction as from the ATNF catalogue results in an average number of detectable sources $\approx 3$. While in absolute terms this is a lower number than for normal neutron stars, it represents a much higher fraction of the population. The fraction is higher by a factor of about 50.

\section{Discussion} 
\label{sec:conclusions}

\subsection{Novelty of approach}

The detectability of continuous gravitational waves depends on the frequency and amplitude of the signal. In turn these quantities are entangled with parameters of the source which are not independent of one another. For instance, the gravitational wave amplitude depends on the ellipticity, spin frequency and on the distance of the source. The frequency depends on the spin frequency at birth, on the age of the object and on the parameters involved in the energy braking mechanisms, e.g. magnetic field and ellipticity. The ellipticity may depend on the magnetic field. Distance and age are not completely independent quantities. 

Several papers in the past two decades have studied the prospects for detection of continuous gravitational waves by Galactic neutron stars \citep{Palomba2005, KnispelAllen2008, Wade2012, Cieslar2021, Soldateschi2021,Lasky2015uia, Woan2018, Reed2021}, but to the best of our knowledge, in no prior detectability analysis, all the effects mentioned above have been taken into account together {\it {consistently}}. 

The first pioneering works of \cite{Palomba2005} and \cite{KnispelAllen2008} consider \emph{gravitars} only, i.e. neutron stars that are loosing rotational energy solely due to gravitational radiation. The gravitar scenario simplifies calculations and is an intriguing one to explore, but it is not clear that such population exists. 

\cite{Wade2012} are the first who take into account magnetised neutron stars, although they consider unphysical populations where each neutron star has the same magnetic field and ellipticity values. 

\cite{Cieslar2021} make a detectability assessment of a population of non-axisymmetric neutron stars all born with the same ellipticity that decays exponentially with time. The distribution of sky positions and frequency is based on a single synthetic population evolved neglecting the gravitational wave spin-down contribution. 

\cite{Soldateschi2021} assess the detectability of continuous waves emitted by a synthetic population of Galactic neutron stars (both normal and recycled) whose deformation is caused by their magnetic field \citep{Soldateschi2021-2}. They do not model the spin evolution but rather sample the spin frequency from the ATNF catalogue and randomly assign magnetic fields.
In doing so, the correlation between magnetic field and spin frequency is lost. This explains why they find currently detectable magnetically deformed sources, which indeed in their case correspond to millisecond pulsars. Current observational results -- no detections -- are in tension with these predictions. 
 
\cite{Reed2021} consider continuous gravitational wave emission from a Galactic population of neutron stars, all equally deformed. They study what fraction of the Galactic population is probed by different searches, as a function of the assumed ellipticity value. Their results characterize the significance of the ellipticity constraints of observational results, and in turn can be used to choose the target parameter space for future searches. The focus of their work is quite different from the one investigated here.

In this paper we produce a synthetic neutron star population, consistently evolved {\it{since birth}} in frequency and position, based on initial positions, kicks, spins, ellipticities and magnetic field values. The age and position of each newly born neutron star is based on the remnant seeding it. 

This is the first study that simulates the remnant population seeding the neutron stars. This is in principle important in order to model dependencies of the remnant mass, position, age with those of the newly born neutron star such as the birth spin frequency, initial kick, magnetic field.  As we discuss in Section~\ref{sec:caveats} in this first study we made a number of simplifying assumptions. Nevertheless we stress here the value of an ``ab initio" framework, because it allows to naturally fold-in more realistic models as they become available, and improve the reliability of the predictions. 

Whereas previous studies have by and large adopted a single model, we consider two broad categories of models: agnostic (A) and empirical (E). The agnostic models use uninformed priors on the broadest physically motivated parameter range possible. The empirical models use more informative priors, reasonably well-accepted in the astrophysical community. From the different detectability profiles that stem from these different populations, we learn about the population parameters that mostly affect the detectability, about the range of detection probability and about what we can learn from non-detections. 

Finally, a number of studies (e.g. \cite{Cieslar2021} and \cite{Soldateschi2021}) adopt a detectability criteria ($h_0^{min.det.}=11.4\sqrt{{S_n}\over{T_{obs}}}$, where $S_n$ is the detector noise and $T_{obs}$ is the observation time) that does not apply to large surveys and overestimates the search sensitivity.
We use a robust detection criterium, based on the {\it{measured}} sensitivity of broad surveys. The minimum detectable intrinsic strain amplitude is $h_0^{min.det.} = {{\sqrt{S_n}}\over {\cal{D}}}$, where $\cal{D}$ is the sensitivity depth of the search \citep{Behnke2015, Dreissigacker2018} and for the isolated-neutron stars surveys considered here ${\cal{D}}\approx 55~ {[1/\sqrt{\textrm{Hz}}]}$. 

\subsection{Results}

We find that normal neutron stars that might be detectable in the foreseeable future must have an ellipticity that is not solely due to magnetic field deformations. In fact whereas large magnetic fields can produce large deformations, they are also responsible for the fast spin-down of the frequency to values too small for a signal to be detectable. For instance, a magnetic field of $10^{14}$ G can source an ellipticity $\approx 10^{-6}$, but yields a spin-down time of $\approx$ 10 years, which makes these sources extremely rare, and practically impossible to find within a kpc of Earth (which is the reach of current detectors for that ellipticity).

To get sufficiently high ellipticities and at the same time avoid fast spin-downs, the poloidal magnetic field component should be much smaller than the toroidal component, i.e. the value of $\Lambda$ should be very small (see Equation~\ref{eq:mastrano}). If we perform our simulations progressively decreasing $\Lambda$, we find that in order to have a 10\% chance of detection\footnote{This estimation was done maintaining $\varepsilon_{max} = 10^{-5}$ as the maximum possible ellipticity.}, $\Lambda \lesssim 10^{-6}$. The stability of such strongly toroidal-dominated configurations is questioned in a number of studies \citep{Lander2009, Ciolfi2010, Lasky2012, Ciolfi2013, Lander2021}.

Detectable objects with not purely magnetic deformations ($\textrm{A2}$ models) have ellipticity values greater than $10^{-7}$ and magnetic fields between $\approx 10^{10}$ and $\approx 10^{12} \, \textrm{G}$. This is a rather narrow range of magnetic field and ellipticity values: at higher magnetic fields the spin-down age decreases making objects spinning in-band more and more rare. The ellipticity further down-selects on this sample, based on the reach of the detectors at that ellipticity value. 
We provide an empirical expression for the expected number of objects for different intervals of $B,\varepsilon$ in this range (Equation~\ref{eq:n_of_B_eps}).
\par\noindent
It is not settled whether values of the ellipticity of the order of $\approx 10^{-6}$, necessary for a detection at current sensitivities, are actually possible. \cite{Gittins2021-1} and \cite{Gittins2021-2} have proposed a new way to assess the maximally sustainable crustal strain in neutron stars. Their most optimistic results generally predict maximum ellipticities of about $\approx 5 \cdot 10^{-7}$ (see Table 1 of \cite{Gittins2021-1}), more than an order of magnitude smaller than the value adopted here at $10^{-5}$ for the largest ellipticity (see Equations~\ref{eq:scenarioOneEpsilon} and \ref{eq:epsilonScenarioTwoLimits}). If we set $5 \cdot 10^{-7}$ as the maximum ellipticity of our synthetic neutron stars, and a log-uniform distribution down to a value determined by the magnetic field (namely Model 2 ellipticities), the outcomes of our simulations change dramatically: the bulk of the loudest signals generally lay about an order of magnitude below the current best upper limits, with only three currently detectable signals out of $100$ realisations from model $\textrm{A2}_{\textrm{high}}$ and only two from model $\textrm{A2}_{\textrm{low}}$. This means a reduction in total number of currently detectable sources by a factor between $40$ and $57$. We however point out that \cite{2022MNRAS.517.5610M} have recently shown that so small ellipticities are not a necessary consequence of the \cite{Gittins2021-1} approach.

The number of detectable sources increases by more than two orders of magnitude for searches on data from 3rd-generation detectors, result in line with what found by \cite{Reed2021}. The increased low-frequency performance allows to intercept a plentiful population of weak sources spinning below 20 Hz. If 3rd-generation detectors do not detect a continuous gravitational signal, this excludes a maximum ellipticity at the $10^{-5}$ level, at least for the very broad $A$ models.

Generally, magnetic field decay increases the chances of detection: for the 3rd-generation detectors we also find that sources drawn from the ``empirical Model" populations of Section~\ref{sec:empirical} begin to become detectable, with model $\textrm{E2}_{\textrm{unif}}$ yielding the most optimistic predictions, with an average of $5.2$ detectable sources. Consistently with this, also \citeauthor{Cieslar2021} -- whose model includes magnetic field decay -- find detectable signals, but overestimate the sources by a factor $\approx$ 5, because their spin evolution neglects the gravitational-wave spin-down, which instead contributes to pushing a significant number of sources out of band (see Section~\ref{sec:models_E2}).

Computational cost might be saved in all-sky surveys by restricting the searches to $\pm 15^\circ$ of the galactic plane and by limiting the frequency derivative range as a function of frequency as discussed in Section~\ref{sec:phase_param} -- presently the only scenarios that produce detectable signals favour the frequency range $[60-550]$ Hz. To what extent such savings could be usefully re-invested yielding a more sensitive search, needs to be evaluated in the context of an optimisation scheme such as that designed for targeted searches \citep{Ming2016, Ming2018}. Our synthetic populations provide a key element for such studies.

Our simple recycled neutron star model predicts that an upcoming detection from these stars is very unlikely, merely due to the much lower number of objects compared to normal neutron stars. However the fraction of objects in-band is much higher and this yields promising prospects for the detectability by 3rd-generation detectors. We find that the expected number of detectable sources by 3rd-generation detectors lays between 0.2 and 6, depending on the ratio of isolated neutron stars to neutron stars in binary systems. Assuming the relative abundance of the two sub-populations the same as the currently observed one in MSPs, the number of detectable sources is $\approx 3$. We stress however that our analysis on recycled neutron stars constitutes an early stage investigation, and contains a number of simplifying assumptions that are discussed in the next Section. 

\subsection{Caveats}
\label{sec:caveats}

\subsubsection{Remnant population}

We assume a single exponential star formation history, spatially invariant across the Galaxy. We neglect the lifespan of massive stars and assume the same fraction of the entire stellar mass to be neutron stars, for progenitors older than about 4 Myr (i.e. the onset age for a stellar population to start producing neutron stars). 
This will overestimate the amount of neutron stars younger than approximately 30 Myr. The impact on the detectability might be significant -- maybe reducing the detection probability even by a factor of 10 -- because most of the detectable objects are younger than $10^7$ years.

We assume a relatively wide mass range for obtaining neutron star remnants. The upper mass limit may decrease if -- for example -- we assume initial rotational velocity in the progenitor.

\subsubsection{Normal neutron star population}
Our modelling of the magnetic field decay is independent of the value of the field at birth. This might constitute an oversimplification -- in fact \cite{Gullon2014} find a correlation between the magnitude of the field at birth and the decay time-scale, with bigger fields decaying much faster than smaller ones. As shown in Figure \ref{fig:decay_test}, most of our future detectable sources in model E2 are born with $B \lesssim 10^{12}\textrm{G}$ and at the present time have fields about 4 orders of magnitudes smaller. It is not clear how a decay time-scale dependent on the magnetic field at birth would impact our results. It may bring in band high-magnetic field sources, whose field would decrease faster than in our models. Conversely, it may push out of the band sources with $B \lesssim 10^{12}\textrm{G}$ that in our models are detectable, due to the slower decay.

\subsubsection{Recycled neutron star population}
We have assumed a constant birthrate of 1 every 200,000 years. The chosen value is important as it directly determines the total number of fast-spinning neutron stars, on which the number of detectable sources directly depends. Since the \emph{re-birth} of a recycled neutron star is not followed by any recordable event, there are no direct measurements of the birthrate, and all predictions stem from population synthesis simulations, and span three orders of magnitude. We follow \cite{Story2007} and consider 60,000 objects, but other studies predict as little as $10^3$ stars \citep{2003ApJ...597.1036P} or as many as $10^6$ \citep{Zhu2015}.

Our analysis of recycled neutron stars ignores accreting systems, which are very interesting as the accretion process may provide a natural source of asymmetry. It has in fact long been proposed that gravitational wave emission could provide the torque-balancing mechanism that explains why no accreting neutron star is spinning anywhere close to the maximum possible spin rate \citep{1984ApJ...278..345W,Patruno_2017}. In this case it can be argued that the gravitational wave amplitude grows with the mass accretion rate, and hence with the luminosity from the accretion process. This is what makes very bright accreting objects like Sco X-1 particularly interesting \citep{Zhang:2020rph}.  
On the other hand, bright objects are easier to observe in the EM domain, and one could argue that the determination of their properties should more usefully rely on EM observations and that the case for population studies like this is less compelling for these systems. 

\subsection{Prospects/Conclusions}

We have presented results from a broad ``ab initio" study of the detectability of continuous gravitational waves emitted by fast rotating neutron stars. This is the first study of this kind.

Our predictions are consistent with the null detection results of the latest all-sky searches. Presently a detection is not excluded but it is limited to stars with deformations that are not just due to the magnetic field (A2 models) and even in this case, it is far from ``guaranteed" (Table~\ref{tab:detect}). In fact the low number of detectable sources means that a change of a factor of two in, say, the size of the progenitor population -- which could very easily come about -- could produce significant changes in the chances of observing the first signal. With detectors just a factor of two more sensitive, the situation changes substantially, making the prospects of detection rather more robust, at least for our agnostic models (Table~\ref{tab:bands}). 

One way to increase the sensitivity is to limit the searched parameter space. Our results indicate that at the present time the detection probability would not be significantly impacted by restricting the search to $\pm 15^\circ $ of the Galactic plane and to limit the frequency range below 600 Hz. We stress again that with low numbers of expected detectable sources, decisions to limit the search space must be carefully evaluated against the sensitivity gains that such savings produce.

The main two enhancement that we foresee concern the remnant population and the modelling of the recycled neutron star population:

\begin{itemize}
\item In future work we plan to trace the formation of neutron stars in young stellar progenitor populations and to consider different star formation histories within our Galaxy, enlightened by recent massive stellar spectroscopic surveys \citep{lian2020a,lian2020b,lian2018a,spitoni2021}.
\item The outcome of the recycling is intimately coupled with the binary parameters, which we have ignored. To consider the recycling process is a project of its own, but we see in it great potential, in providing guidance on how to best search the orbital parameter space. 
\end{itemize}

Thanks to the upcoming pulsar surveys, the number of known neutron stars is expected to grow in the course of this decade to over 20,000 \citep{Smits_2009}. While this will probably not significantly advance our understanding of the degree of deformation of neutron stars, it will shed light on the evolution of neutron stars and on their parameters, such as their spin and spatial distributions, age and magnetic field. Feeding into studies like the one presented here, this information will allow to make more reliable predictions on the parameters of detectable continuous gravitational waves, and will guide the observational surveys. Once signals are detected, these studies will enable inferences on the properties of the underlying population -- including properties that electromagnetic observations are completely blind to. 

\begin{acknowledgements}
The authors are grateful to Bernard F. Schutz for insightful feedback and discussions. D.T. is supported by the Sherman Fairchild Postdoctoral Fellowship at Caltech.
\end{acknowledgements}

\appendix
\numberwithin{equation}{section}
\section{Numerical integration}
\label{sec:num_integ}
\subsection{A models}
In the A models the magnetic fields are independent of time. Equation~\ref{eq:combsd} can be analytically integrated and we obtain Equation~\ref{eq:combsdintegr}. Unfortunately, the latter equation cannot be inverted to give $\nu(t)$. We follow \citeauthor{Wade2012} and find the frequency at present time as one of the zeroes of Equation~\ref{eq:combsdintegr} consistent with a monotonic spin-down. We do this using the root-finding method {\tt{brentq}} from the {\tt{scipy}} library.

\subsection{E models}
For these models the magnetic fields have the time-dependence defined in Equation~\ref{eq:decay} and it is not possible to analytically integrate Equation~\ref{eq:combsd}. 

We proceed as follows.
Unlike the time-independent magnetic field case, $\gamma$, defined in Equation~\ref{eq:upperGamma}, is a monotonically increasing function of time through $B(t)$. 
If Equation~\ref{eq:upperGamma} is satisfied at birth time but not at present time, then it means that there exists a time $t^*$ such that 
\begin{equation}
\label{eq:gammatstar}
\gamma(t^*) = 10^{-8}\,s^2.
\end{equation} 
We recall that our condition for pure magnetic dipole emission is Equation~\ref{eq:upperGamma}, $\gamma < 10^{-8}\,s^2$. So Equation~\ref{eq:gammatstar}, tells us that the condition for pure magneto-dipole spin-down is satisfied until $t^*$, and for $t> t^*$ the magnetic field is so small that we cannot anymore ignore the gravitational wave spin-down contribution.
We thus integrate analytically, approximating the spin-down to be purely magneto-dipolar until $t^*$, and then integrate numerically up to the present time.
If Equation~\ref{eq:upperGamma} is not valid at birth, we integrate numerically for the entire age of the object.
We use the {\tt{odeint}} function of the {\tt{scipy}} library limiting the maximum number of steps per integration to 10,000.

\section{Intrinsic detection odds as a function of birth spin-frequency}
\label{sec:factor_7}
Starting from the results of our synthetic population, we want to compare the ``intrinsic" chances of detection for sources whose birth spin frequency belongs to the two different bands, $\mathcal{B}_1 = [2, 300]\,\textrm{Hz}$ and $\mathcal{B}_2 = [300, 1200]\,\textrm{Hz}$, i.e. factoring out the fact that the two bands have a different size. We consider the results from model $\textrm{A2}_{\textrm{high}}$.

In the hundred realisations performed, we find $362$ currently detectable sources. Of these, $124$ ($\approx 34\%$ of $362$) have a birth spin frequency $\nu_0 \in \mathcal{B}_1$, while the remaining $238$ ($\approx 66\%$ of $362$) have $\nu_0 \in \mathcal{B}_2$. In the $\textrm{A2}_{\textrm{high}}$ model, birth spin frequencies are distributed log-uniformly between 2 and 1200 Hz; that means that $\approx 78\%$ of the neutron star population is born with $\nu_0 \in \mathcal{B}_1$ and the remaining $\approx 22\%$ with $\nu_0 \in \mathcal{B}_2$ (the two bands have indeed different sizes). If $N^{TOT}$ is the total number of sources in the hundred realisations performed, there are $\approx 0.78 \cdot N^{TOT}$ sources born with $\nu_0 \in \mathcal{B}_1$ and $\approx 0.22 \cdot N^{TOT}$ born with $\nu_0 \in \mathcal{B}_2$. The two fractions 
\begin{equation}
\label{eq:odds}
\mathcal{F}^{det}_{\mathcal{B}_1} = \frac{124}{0.78 \cdot N^{TOT}}{\textrm{~~and~~}}  \mathcal{F}^{det}_{\mathcal{B}_2} = \frac{238}{0.22 \cdot N^{TOT}}
\end{equation}
represent the chances that a source be detectable \emph{given} that it was born with $\nu_0 \in \mathcal{B}_1$ and $\nu_0 \in \mathcal{B}_2$, respectively, regardless of the size of the two bands.
In practise, for every source born with $\nu_0 \in \mathcal{B}_1$ ($\nu_0 \in \mathcal{B}_2$), a fraction $\mathcal{F}^{det}_{\mathcal{B}_1}$ ($\mathcal{F}^{det}_{\mathcal{B}_2}$) are detectable.
Finally, the ratio
\begin{equation}
\label{eq:F_ratios}
\Lambda = \frac{ \mathcal{F}^{det}_{\mathcal{B}_2} }{ \mathcal{F}^{det}_{\mathcal{B}_1} }
\end{equation}
gives the ``intrinsic" detection odds for a signal from a source born with $\nu_0 \in \mathcal{B}_2$ against those from a source born with $\nu_0 \in \mathcal{B}_1$. Plugging \ref{eq:odds} in \ref{eq:F_ratios} we obtain $\Lambda \approx 6.8$.

\bibliography{bibliography}{}
\bibliographystyle{aasjournal}



\end{document}